\documentclass[journal,twoside,web]{ieeecolor}

\usepackage{generic}
\usepackage{cite}
\usepackage{amsmath,amssymb,amsfonts}
\usepackage{algorithmic}
\usepackage{graphicx}
\usepackage{textcomp}
\usepackage{color}
\usepackage{multirow}
\usepackage{hyperref}
\hypersetup{
	colorlinks=true,
	linkcolor=blue,
	citecolor=blue,
	urlcolor=blue,
}

\let\proof\relax  
\usepackage{amsthm}
\newtheoremstyle{mytheoremstyle} 
    {.3cm}      
    {.3cm}      
    {}          
    {}          
    {\bfseries}  
    {:}         
    {.5em}      
    {}          
\theoremstyle{mytheoremstyle}

\renewcommand{\qedsymbol}{$\blacksquare$}
\newtheorem{thm}{Theorem}

\newtheorem{prop}{Proposition}

\theoremstyle{definition}
\newtheorem{defn}{Definition}
\newtheorem{exmp}{Example}
\newtheorem{rem}{Remark}
\newtheorem{assump}{Assumption}

\newcommand{\dhdx}{\nabla h}
\newcommand{\dHdx}{\nabla_{x} H}
\newcommand{\dHde}{\nabla_{e} H}
 
\begin{document}

\title{Safety-Critical Control with Input Delay \\ in Dynamic Environment}
\author{Tamas G. Molnar, \IEEEmembership{Member, IEEE}, Adam K. Kiss, \\
Aaron D. Ames, \IEEEmembership{Fellow, IEEE}, and G\'abor Orosz, \IEEEmembership{Senior Member, IEEE}
\thanks{This research is supported in part by the National Science Foundation (CPS Award \#1932091), Aerovironment and Dow (\#227027AT), and supported by the NRDI Fund (TKP2020 IES, Grant No. BME-IE-MIFM and TKP2020 NC, Grant No. BME-NC).}
\thanks{Tam\'as G. Moln\'ar and Aaron D. Ames are with the Department of Mechanical and Civil Engineering, California Institute of Technology, Pasadena, CA 91125, USA, (e-mail: tmolnar@caltech.edu, ames@caltech.edu).}%
\thanks{Adam K. Kiss is with the MTA-BME Lend{\"{u}}let Machine Tool Vibration Research Group, Department of Applied Mechanics, Budapest University of Technology and Economics, Budapest 1111, Hungary (e-mail: kiss\_a@mm.bme.hu).}%
\thanks{G\'abor Orosz is with the Department of Mechanical Engineering and with the Department of Civil and Environmental Engineering, University of Michigan, Ann Arbor, MI 48109, USA (e-mail: orosz@umich.edu).}%
}

\maketitle
\thispagestyle{empty}
\pagestyle{empty}

\begin{abstract}
Endowing nonlinear systems with safe behavior is increasingly important in modern control.
This task is particularly challenging for real-life control systems that operate in dynamically changing environments.
This paper develops a framework for safety-critical control in dynamic environments, by establishing the notion of {\em environmental control barrier functions (ECBFs)}.
Importantly, the framework is able to guarantee safety even in the presence of {\em input delay}, by accounting for the evolution of the environment during the delayed response of the system.
The underlying control synthesis relies on predicting the future state of the system and the environment over the delay interval, with robust safety guarantees against prediction errors.
The efficacy of the proposed method is demonstrated by a simple adaptive cruise control problem and a more complex robotics application on a Segway platform.
\end{abstract}

\begin{IEEEkeywords}
Delay systems, Dynamic environment, Predictive control, Robust control, Safety-critical control
\end{IEEEkeywords}

\section{Introduction}
\label{sec:intro}

\begin{figure}[!t]
\centerline{\includegraphics[width=\columnwidth]{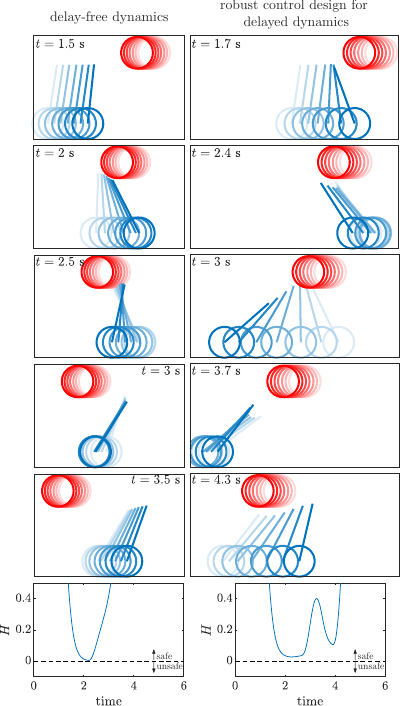}}
\caption{
The proposed safety-critical control framework in high-fidelity simulation of a Segway.
The Segway safely avoids a moving obstacle, even when the obstacle's future position is unknown and there is input delay in the control loop.
This is accomplished via the environmental control barrier function (ECBF) plotted at the bottom.
Observe that the Segway uses different strategies in the delay-free and delayed cases.
See video at \url{https://youtu.be/NIImeVnlziM}.
\vspace{-3mm}
}
\label{fig:concept}
\end{figure}

\IEEEPARstart{S}{afety}
is of great importance in many control systems, including a wide spectrum of applications from automated vehicles~\cite{Niletal2016, HeOrosz2018} through robotics~\cite{Teng2021, tordesillas2019faster, Kousik2020, Nubert2020} and multi-robot systems~\cite{Panagou2016, glotfelter2017nonsmooth, Srinivasan2018}, to controlling the spread of infectious diseases~\cite{ames2020safety,Molnar2021lcss}.
Notably, safety is often affected by a dynamic environment that surrounds the control system.
For example, robots must avoid collision with other agents in multi-robot systems~\cite{Falconi2014, Santillo2021},
automated vehicles must drive safely amongst other road users~\cite{schwarting2018planning}, and robotic manipulators must collaborate safely with their human operator~\cite{Zanchettin2016, Landi2019, Singletary2019}.

Strict safety requirements call for theoretical safety guarantees and provably safe controllers. 
Thus, control synthesis must take into account how the control system interacts with its environment, and it must ensure that environment's evolution does not lead to safety violations.
As such, dynamic environments pose a major challenge for safety-critical control.

An important element of this challenge is that the response time of control systems may be commensurate with how fast the environment changes.
Response times include sensory, feedback and actuation delays that arise in practice~\cite{Stepan1989}.
The magnitude of the delay depends on the application: it is milliseconds in robotic systems~\cite{Andersen2015},
a few tenths of a second in automated vehicles~\cite{Ji2021}
and days in epidemiological models~\cite{casella2021covid19}.
Delays significantly impact safety in dynamic environments, since by the time the control system responds, the environment may change and safety may be compromised.
To overcome this danger, one must consider how the dynamic -- and often uncertain -- environment evolves over the delay period, which yields a major challenge in designing provably safe controllers.
This paper addresses this problem by establishing a framework for safety-critical control that takes dynamic environments and time delays into account explicitly.


\vspace{-3pt}
\subsection{State of the Art}

Formally, safety is often framed as a set invariance problem by requiring the state of the system to evolve within a safe set for all time.
The theory of {\em control barrier functions (CBFs)} provides an elegant solution to achieve this goal~\cite{AmesXuGriTab2017}. 
While this theory delivers formal safety guarantees, one shall secure these guarantees in dynamic environments during practical implementation.
Several works have built on CBFs to transfer safety-critical controllers from theory to practice, by providing robustness against disturbances~\cite{jankovic2018robust, ames2019issf, choi2021robust, Zheng2021, Alan2022}, measurement uncertainty~\cite{takano2018robust, Clark2021, Dean2021} and model mismatches~\cite{wang2018safe, choi2020reinforcement}.
Safety in dynamic environments were addressed by \cite{Zhu2019} and \cite{Luo2020} in the context of collision avoidance in multi-agent systems by incorporating chance constraints into model predictive control and probabilistic safety barrier certificates, respectively.
Furthermore,~\cite{Igarashi2019, Tezuka2020} discussed human assist control, in which response to changing environments was handled via time-varying CBFs, including adaptivity to unknown environment parameters and robustness to disturbances.

The safety of time delay systems has been attracting increasing attention.
The safety of continuous-time systems with state delay was established by safety functionals in~\cite{OroAme2019, Kiss2021}, which were extended to control barrier functionals in~\cite{liu2021safety, Kiss2022}.
Discrete-time control systems with input delay were studied in~\cite{necmiye2020} for linear and in~\cite{singletary2020control} for nonlinear dynamics.
Linear systems with input delays were addressed in continuous time in~\cite{Jankovic2018,abel2019constrained} via control barrier and Lyapunov functions.
Safety-critical control of continuous-time nonlinear systems with measurement delays was tackled in our works~\cite{ames2020safety,Molnar2021lcss} in an application to controlling the spread of COVID-19.
These papers leveraged predictor feedback~\cite{Krsticbook2008,BekKrs2013,Karafyllis2017,michiels2007stability} to compensate the delay by predicting the future evolution of the system.

Remarkably,~\cite{singletary2020control} also relied on predictor feedback to compensate input delay.
The underlying theory was established in discrete time by assuming that the system's evolution is predicted accurately.
As opposed, here we consider continuous-time systems and address prediction errors.
Parallel to our work,~\cite{abel2020constrained} proposed predictor feedback in continuous time for compensating multiple input delays in safety-critical control, wherein input channels with shorter delays were used to keep the system safe until longer delays were compensated.
\cite{Abel2021} modified the predictor to compensate time-varying input delay and achieve safety.
\cite{Molnar2022tds} endowed safety-critical predictor feedback controllers with robustness against disturbances.
Yet, these works have not addressed
safety in dynamic environments that evolve independently of the control input.
This paper intends to fill this gap and tackle the challenges arising from the combination of dynamic environments and delays.

In this paper, we {\em explicitly} involve {\em dynamically changing environments} into the framework for safety-critical control, in order to handle uncertain environments with worst-case safety guarantees in a deterministic fashion.
Importantly, our framework also allows us to compensate the effect of input delay, which was not addressed in the literature.

\subsection{Contributions}

Here we build on~\cite{ames2020safety,Molnar2021lcss} to establish the theory of safety-critical control for nonlinear continuous-time systems with input delay, operating in dynamically changing environments.
Our contributions are threefold:
\begin{itemize}
\item[1.]
We establish the notion of {\em environmental control barrier functions (ECBFs)} for delay-free systems to explicitly address scenarios in which safety is affected by a dynamic environment.
This notion is particularly useful when the dynamics of the environment are inherently more uncertain than those of the control system.
\item[2.]
We connect the theories of CBFs and predictor feedback by developing the notions of CBFs and ECBFs for systems with input delay and synthesizing safety-critical controllers via predictor feedback.
Predictors require special care in dynamic environments as the environment's future cannot be predicted accurately.
Thus, we make controllers robust against prediction errors.
\item[3.]
We demonstrate the efficacy of this framework on real-life engineering systems where time delays and dynamic environments both occur, through the examples of adaptive cruise control and obstacle avoidance with a Segway.
\end{itemize}

Figure~\ref{fig:concept} illustrates a sample of these results.
A Segway is controlled to safely avoid a moving obstacle via the proposed ECBFs in high-fidelity simulation.
Without delay in its control loop (left), the Segway pitches backwards to go under the obstacle.
With input delay (right), the Segway approaches the obstacle, then moves in reverse to make space, and pitches forward to go under it.
Remarkably, these safe behaviors emerge from the ECBF automatically, which handles reactive planning in a holistic fashion.

The paper is structured as follows.
Section~\ref{sec:CBF} revisits CBFs for delay-free systems.
Section~\ref{sec:environment} addresses safety in dynamic environments by introducing ECBFs.
Section~\ref{sec:delay} extends CBFs and ECBFs to systems with input delays, and discusses safety-critical control via predictor feedback with robustness against prediction errors.
In these sections, adaptive cruise control is used as illustrative example, whereas Section~\ref{sec:segway} demonstrates the safety-critical control of a Segway by numerical simulations. 
We conclude our work in Section~\ref{sec:concl}.

\section{Preliminaries to Safety-Critical Control}
\label{sec:CBF}

Consider a control-affine system with state ${x(t) \in X \subseteq \mathbb{R}^{n}}$ and control input ${u(t) \in U \subseteq \mathbb{R}^{m}}$:
\begin{equation}
\dot{x}=f(x) + g(x) u,
\label{eq:system}
\end{equation}
where ${f: X \to \mathbb{R}^n}$ and ${g: X \to \mathbb{R}^{n \times m}}$ are locally Lipschitz continuous on $X$.
Let ${x(0)=x_{0} \in X}$ be the initial condition.
When the input ${u=k(x)}$ is given by a locally Lipschitz continuous controller ${k: X \to U}$, system~(\ref{eq:system}) has a unique solution over a time interval ${t \in I(x_{0})}$.
For simplicity, we assume ${I(x_{0}) = [0,\infty)}$, i.e., the solution exists for all ${t \geq 0}$.

We consider the system safe if its state is contained within a {\em safe set} ${S \subset X}$ for all time.
Accordingly, we frame safety-critical control as rendering set $S$ forward invariant under dynamics~(\ref{eq:system}): the controller needs to ensure for all ${x_{0} \in S}$ that ${x(t) \in S}$, ${\forall t \geq 0}$.
Specifically, we define $S$ as the 0-superlevel set of a continuously differentiable function ${h: X \to \mathbb{R}}$:
\begin{equation}
S=\{x \in X: h(x) \geq 0 \},
\label{eq:safeset}
\end{equation}
where the selection of $h$ is application-driven.

\subsection{Control Barrier Functions}

We ensure the forward invariance of the safe set $S$ by the framework of {\em control barrier functions (CBFs)}.
First, we briefly revisit the main result in~\cite{AmesXuGriTab2017} that establishes the definition of CBFs and the theoretical safety guarantees.
We use the notation ${\| . \|}$ for Euclidean norm, and we call a function ${\alpha: (-a,b) \to \mathbb{R}}$, ${a,b > 0}$ as extended class $\mathcal{K}$ function, if it is continuous, strictly monotonically increasing and ${\alpha(0)=0}$.

\begin{defn}\label{defn:CBF}
Function $h$ is a control barrier function (CBF) for~(\ref{eq:system}) if there exists an extended class $\mathcal{K}$ function $\alpha$ such that for all ${x \in S}$:
\begin{equation}
\sup_{u \in U} \dot{h}(x,u) > - \alpha(h(x)),
\label{eq:CBF_condition}
\end{equation}
where:
\begin{equation}
\dot{h}(x,u) = \dhdx(x) (f(x) + g(x) u)
\label{eq:hdot}
\end{equation}
is the derivative of $h$ along system~(\ref{eq:system}).
\end{defn}
Note that $\sup$ becomes $\max$ if $U$ is compact.
With the CBF definition,~\cite{AmesXuGriTab2017} establishes formal safety guarantees as follows.

\begin{thm}[\!\!\cite{AmesXuGriTab2017}]\label{thm:safety}
\textit{
If $h$ is a CBF for~(\ref{eq:system}), then any locally Lipschitz continuous controller ${u=k(x)}$ satisfying:
\begin{equation}
\dot{h}(x,u) \geq - \alpha(h(x)),
\label{eq:safety_condition}
\end{equation}
${\forall x \in S}$ renders $S$ forward invariant (safe), i.e., it ensures ${x_{0} \in S \Rightarrow x(t) \in S}$, ${\forall t \geq 0}$.
}
\end{thm}

The proof can be found in~\cite{AmesXuGriTab2017}, and further technical details with discussion about the selection of $\alpha$ are in~\cite{Konda2021}.
Throughout the paper, we use variants of the safety condition~(\ref{eq:safety_condition}).

\begin{rem}
Condition~(\ref{eq:safety_condition}) is often used in the context of optimization-based controllers~\cite{AmesXuGriTab2017}.
Given a control input ${u_{\rm d}=k_{\rm d}(x)}$ by a desired controller ${k_{\rm d}: X \to U}$, one can modify this input in a minimally invasive fashion to guarantee safety by solving the following quadratic program (QP):
\begin{align}
\begin{split}
k(x) = \mathrm{arg}\hspace{-.1cm}\min_{\hspace{-.2cm} u \in U} & \quad \|u - k_{\rm d}(x)\|^2  \\
\mathrm{s.t.} & \quad \dot{h}(x,u) \geq - \alpha(h(x)).
\end{split}
\label{eq:QP}
\end{align}
This defines the control law ${u = k(x)}$ {\em implicitly}.
The feasibility of this QP is guaranteed by the definition of CBFs (Definition~\ref{defn:CBF}).
However, verifying that a given $h$ is indeed a CBF is nontrivial when there are input constraints (${U \subset \mathbb{R}^{m}}$).
An approach to overcome input constraints is the backup set method~\cite{Singletary2019}, that relies on the forward integration of the dynamics similar to predictor feedback presented in Section~\ref{sec:delay}.
Otherwise, without input bounds (${U = \mathbb{R}^{m}}$) feasibility guarantees can be proven, and the solution to QP~(\ref{eq:QP}) can even be expressed {\em explicitly} based on the KKT conditions~\cite{Boyd2004} as
${k(x) = k_{\rm d}(x)}$ if ${\dhdx(x) g(x) = 0}$ and:
\begin{align}
\begin{split}
k(x) & = k_{\rm d}(x) + \max\{ - \phi_{0}(x), 0 \} \phi_{1}^+(x), \\
\phi_{0}(x) & = \dhdx(x) (f(x) + g(x) k_{\rm d}(x)) + \alpha(h(x)), \\
\phi_{1}(x) & = \dhdx(x) g(x),
\end{split}
\label{eq:QP_solution}
\end{align}
if ${\dhdx(x) g(x) \neq 0}$, where ${\phi_{1}^+(x) = \phi_{1}^\top(x) / ( \phi_{1}(x) \phi_{1}^\top(x))}$ is the right pseudoinverse of $\phi_{1}(x)$.
The derivation of~(\ref{eq:QP_solution}) is given in Appendix~\ref{sec:appdx_KKT}.
Note that if ${\dhdx(x) g(x) \neq 0}$, ${\forall x \in S}$, it is often referred to as $h$ has {\em relative degree}~$1$ (i.e., the first derivative of $h$ with respect to time is affected by $u$).
For higher relative degrees (when a higher derivative of $h$ is affected by $u$), there exist systematic methods to construct CBFs from $h$ and guarantee safety; see~\cite{Nguyen2016, Xiao2019, sarkar2020highrelative, Wang2020LearningCB} for details.
An example for such extension is given later in Section~\ref{sec:segway}.
\end{rem}

\section{Safety in Dynamic Environment}
\label{sec:environment}

So far we related safety to the state $x(t)$ of the system.
Often safety is also affected by the state of the environment, which we characterize by ${e(t) \in E \subseteq \mathbb{R}^{l}}$, where $e$ is a continuously differentiable function of time with ${\dot{e}(t) \in \mathcal{E} \subseteq \mathbb{R}^{l}}$ and ${e(0)=e_{0} \in E}$.
This leads to an {\em environmental safe set} $S_{\rm e}$:
\begin{equation}
S_{\rm e}=\{(x,e) \in X \times E: H(x,e) \geq 0 \},
\label{eq:safeset_environment}
\end{equation}
where ${H: X \times E \to \mathbb{R}}$ is assumed to be continuously differentiable in both arguments.

\subsection{Environmental Control Barrier Functions}

We enforce safety in dynamic environments by introducing the notion of {\em environmental control barrier functions (ECBFs)}.
\begin{defn}\label{defn:CBF_environment}
Function $H$ is an environmental control barrier function (ECBF) for~(\ref{eq:system}) if there exists an extended class $\mathcal{K}$ function $\alpha$ such that for all ${(x,e) \in S_{\rm e}}$ and ${\dot{e} \in \mathcal{E}}$:
\begin{equation}
\sup_{u \in U} \dot{H}(x,e,\dot{e},u) > - \alpha(H(x,e)),
\label{eq:CBF_condition_environment}
\end{equation}
where:
\begin{equation}
\dot{H}(x,e,\dot{e},u) = \dHdx(x,e) (f(x) + g(x) u) + \dHde(x,e) \dot{e}
\label{eq:hdot_environment}
\end{equation}
is the derivative of $H$ along system~(\ref{eq:system}).

\end{defn}

ECBFs are a time-dependent extension of CBFs, wherein the dependence on time is considered through the state $e(t)$ of the environment.
This will facilitate addressing environment uncertainty in Section~\ref{sec:ECBF_robustness}.
The ECBF condition~(\ref{eq:CBF_condition_environment}) is directly related to the CBF condition~(\ref{eq:CBF_condition}), with an additional term in the derivative with respect to time.
Further literature on time-varying CBFs can be found, for example, in~\cite{Igarashi2019, Tezuka2020}.

Via the ECBF, an extension of Theorem~\ref{thm:safety} yields theoretical safety guarantees in dynamic environments, as given below.

\begin{thm}\label{thm:safety_environment}
\textit{
If $H$ is an ECBF for~(\ref{eq:system}), then any locally Lipschitz continuous controller ${u=K(x,e,\dot{e})}$ satisfying:
\begin{equation}
\dot{H}(x,e,\dot{e},u) \geq - \alpha(H(x,e)),
\label{eq:safety_condition_environment}
\end{equation}
${\forall (x,e) \in S_{\rm e}}$ and ${\forall \dot{e} \in \mathcal{E}}$ renders $S_{\rm e}$ forward invariant, i.e., it ensures ${(x_{0},e_{0}) \in S_{\rm e} \Rightarrow (x(t),e(t)) \in S_{\rm e}}$, ${\forall t \geq 0}$.
}
\end{thm}

\proof
(\ref{eq:system}) and its environment form the augmented system:
\begin{equation}
\dot{z} = F(z) + G(z) v,
\end{equation}
with augmented state $z$, input $v$ and dynamics $F$ and $G$ as:
\begin{equation}
z =
\begin{bmatrix}
x \\ e
\end{bmatrix}, \quad
v =
\begin{bmatrix}
u \\ \dot{e}
\end{bmatrix}, \quad
F(z) =
\begin{bmatrix}
f(x) \\ 0
\end{bmatrix}, \quad
G(z) =
\begin{bmatrix}
g(x) \\ I
\end{bmatrix}.
\end{equation}
For this system, function ${H_{z}: X \times E \to \mathbb{R}}$, ${H_{z}(z) = H(x,e)}$ is a CBF, since $H$ is an ECBF.
Based on Theorem~\ref{thm:safety}, safety is guaranteed with respect to the 0-superlevel set of $H_{z}$ by:
\begin{equation}
\dot{H}_{z}(z,v) \geq - \alpha(H_{z}(z)).
\end{equation}
Substituting the definitions of $z$, $v$, $F$, $G$ and $H_{z}$ leads to~(\ref{eq:safety_condition_environment}) and proves the statement in Theorem~\ref{thm:safety_environment}.
\hfill \qedsymbol

\begin{rem}
Theorem~\ref{thm:safety_environment} yields safety-critical controllers of the form ${K: X \times E \times \mathcal{E} \to U}$, ${u = K(x,e,\dot{e})}$ that depend on the environment as well through $e$ and $\dot{e}$.
For example, a controller based on optimization (specifically, a QP) reads:
\begin{align}
\begin{split}
K(x,e,\dot{e}) =
\mathrm{arg}\hspace{-.1cm}\min_{\hspace{-.2cm} u \in U} & \quad \|u - K_{\rm d}(x,e,\dot{e})\|^2  \\
\mathrm{s.t.} & \quad \dot{H}(x,e,\dot{e},u) \geq - \alpha(H(x,e)),
\end{split}
\label{eq:QP_environment}
\end{align}
analogously to~(\ref{eq:QP}), with explicit solution for ${U = \mathbb{R}^{m}}$:
\begin{align}
\begin{split}
K(x,e,\dot{e}) & = K_{\rm d}(x,e,\dot{e}) + \max\{ - \Phi_{0}(x,e,\dot{e}), 0 \} \Phi_{1}^+(x,e), \\
\Phi_{0}(x,e,\dot{e}) & = \dHdx(x,e) (f(x) + g(x) K_{\rm d}(x,e,\dot{e})) \\
& \quad + \dHde(x,e) \dot{e} + \alpha(H(x,e)), \\
\Phi_{1}(x,e) & = \dHdx(x,e) g(x),
\end{split}
\label{eq:QP_solution_environment}
\end{align}
analogously to~(\ref{eq:QP_solution}) if ${\dHdx(x,e) g(x) \neq 0}$.
\end{rem}

\subsection{Robust Safety in Uncertain Environment}
\label{sec:ECBF_robustness}

ECBFs rely on the environment's state $e$ and its derivative $\dot{e}$.
In practice, these quantities are typically estimated with uncertainty.
Thus, now we robustify safety-critical controllers against uncertainties in the environment.
Motivated by the method developed in~\cite{Dean2021}
for handling state uncertainty, we provide robustness based on worst-case uncertainty bounds (i.e., in a deterministic fashion).
For simplicity, we consider no uncertainty in $x$, since the environment is typically associated with more uncertainty than the state of the control system.

Consider that the true environment state $e$ and its derivative $\dot{e}$ are not available, only some estimates $\hat{e}$ and $\hat{\dot{e}}$.
We assume these estimates have known uncertainty bounds $\varepsilon_{e}$ and $\varepsilon_{\dot{e}}$: 
\begin{equation}
\| e - \hat{e} \| \leq \varepsilon_{e}, \quad
\big\| \dot{e} - \hat{\dot{e}} \big\| \leq \varepsilon_{\dot{e}}.
\label{eq:environment_uncertainty_bound}
\end{equation}
While it may be nontrivial to find such error bounds, conservative over-approximations of uncertainty bounds are usually available in practice for many perception, measurement or state estimation algorithms.
As such, the approach proposed below is limited to setups with known uncertainty bounds, since safety is guaranteed by considering the worst-case scenario.

The main idea is to enforce safety through a conservative lower bound on the unknown expression $\dot{H}(x,e,\dot{e},u) + \alpha(H(x,e))$ that must be kept nonnegative per Theorem~\ref{thm:safety_environment}.
The bound uses the known quantities $\hat{e}$ and $\hat{\dot{e}}$ in the form:
\begin{multline}
\dot{H}(x,e,\dot{e},u) + \alpha(H(x,e)) \\
\geq \dot{H}(x,\hat{e},\hat{\dot{e}},u) + \alpha(H(x,\hat{e})) - C(\varepsilon_{e},\varepsilon_{\dot{e}},u) \geq 0.
\end{multline}
This is stated more formally with the specific expression of $C(\varepsilon_{e},\varepsilon_{\dot{e}},u)$ below, after some additional assumptions.

Assume that the following regularity conditions on $H$ hold.
Functions
${\dHdx(x,e)f(x)}$,
${\dHdx(x,e)g(x)}$
and ${\alpha(H(x,e))}$ are Lipschitz continuous in argument $e$ on $S_{\rm e}$ with Lipschitz coefficients $\mathcal{L}_{\nabla H f,e}$, $\mathcal{L}_{\nabla H g,e}$ and $\mathcal{L}_{\alpha \circ H,e}$, whereas ${\dHde(x,e)\dot{e}}$
is Lipschitz continuous in $e$ and $\dot{e}$ on ${S_{\rm e} \times \mathcal{E}}$ with Lipschitz coefficients $\mathcal{L}_{\nabla H \dot{e},e}$ and $\mathcal{L}_{\nabla H \dot{e},\dot{e}}$.
This implies:
\begin{align}
\begin{split}
& {\dHdx f}|_{x,e} - {\dHdx f}|_{x,\hat{e}}
\geq - \mathcal{L}_{\nabla H f,e} \| e - \hat{e} \|, \\
& \big( {\dHdx g}|_{x,e} - {\dHdx g}|_{x,\hat{e}} \big) u
\geq - \mathcal{L}_{\nabla H g,e} \| e - \hat{e} \| \| u \|, \\
& {\dHde \dot{e}}|_{x,e,\dot{e}} - {\dHde \dot{e}}|_{x,\hat{e},\hat{\dot{e}}} \\
& \quad \geq - \mathcal{L}_{\nabla H \dot{e},e} \| e - \hat{e} \| - \mathcal{L}_{\nabla H \dot{e},\dot{e}} \| \dot{e} - \hat{\dot{e}} \|, \\
& {\alpha \circ H}|_{x,e} - {\alpha \circ H}|_{x,\hat{e}}
\geq - \mathcal{L}_{\alpha \circ H,e} \| e - \hat{e} \|.
\end{split}
\label{eq:Lipschitz_bounds_hedot}
\end{align}
This leads to the following sufficient condition for safety.

\begin{prop}\label{prop:mrcbf_environment}
\textit{
If $H$ is an ECBF for~(\ref{eq:system}) and the regularity conditions in~(\ref{eq:Lipschitz_bounds_hedot}) hold, then any locally Lipschitz continuous controller ${u=K(x,\hat{e},\hat{\dot{e}})}$ satisfying:
\begin{equation}
\dot{H}(x,\hat{e},\hat{\dot{e}},u) - C(\varepsilon_{e},\varepsilon_{\dot{e}},u) \geq - \alpha \big( H (x,\hat{e}) \big),
\label{eq:safety_condition_mrcbf_environment}
\end{equation}
with:
\begin{multline}
C(\varepsilon_{e},\varepsilon_{\dot{e}},u) = (\mathcal{L}_{\nabla H f,e} + \mathcal{L}_{\alpha \circ H,e} + \mathcal{L}_{\nabla H \dot{e},e}) \varepsilon_{e} \\
+ \mathcal{L}_{\nabla H \dot{e},\dot{e}} \varepsilon_{\dot{e}} + \mathcal{L}_{\nabla H g,e} \varepsilon_{e} \| u \|,
\label{eq:robustify_environment}
\end{multline}
${\forall (x,\hat{e}) \in S_{\rm e}}$ and ${\forall \hat{\dot{e}} \in \mathcal{E}}$ renders $S_{\rm e}$ forward invariant, i.e., it ensures ${(x_{0},e_{0}) \in S_{\rm e} \Rightarrow (x(t),e(t)) \in S_{\rm e}}$, ${\forall t \geq 0}$.
}
\end{prop}

\proof
The steps of the proof follow those of Theorem~2 in~\cite{Dean2021}: we show that~(\ref{eq:safety_condition_mrcbf_environment}) implies~(\ref{eq:safety_condition_environment}) and we apply Theorem~\ref{thm:safety_environment}.
We relate~(\ref{eq:safety_condition_mrcbf_environment}) to~(\ref{eq:safety_condition_environment}) by introducing the difference between their corresponding terms.
By using~(\ref{eq:hdot_environment}), we get:
\begin{align}
\begin{split}
\dot{H} & (x,e,\dot{e},u) + \alpha(H(x,e)) \\
& = \dot{H}(x,\hat{e},\hat{\dot{e}},u) + \alpha(H(x,\hat{e})) \\
& + {\dHdx f}|_{x,e} - {\dHdx f}|_{x,\hat{e}}
+ \big( {\dHdx g}|_{x,e} - {\dHdx g}|_{x,\hat{e}} \big) u \\
& + {\dHde \dot{e}}|_{x,e,\dot{e}} - {\dHde \dot{e}}|_{x,\hat{e},\hat{\dot{e}}}
+ {\alpha \circ H}|_{x,e} - {\alpha \circ H}|_{x,\hat{e}}.
\end{split}
\end{align}
These differences show up in~(\ref{eq:Lipschitz_bounds_hedot}).
Thus, the regularity conditions~(\ref{eq:Lipschitz_bounds_hedot}) on $H$, the uncertainty bound~(\ref{eq:environment_uncertainty_bound}) and condition~(\ref{eq:safety_condition_mrcbf_environment},~\ref{eq:robustify_environment}) imply~(\ref{eq:safety_condition_environment}), which completes the proof.
\hfill \qedsymbol

Less conservative problem-specific bounds than~(\ref{eq:robustify_environment}) also work as long as they imply~(\ref{eq:safety_condition_environment}).
Furthermore, we highlight that~(\ref{eq:robustify_environment}) involves the term $\| u \|$.
This, when incorporated into an optimization problem like~(\ref{eq:QP}), leads to a second-order cone program (SOCP) rather than a QP if ${\mathcal{L}_{\nabla H g,e} \neq 0}$.

\begin{figure}[!t]
\centerline{\includegraphics[width=\columnwidth]{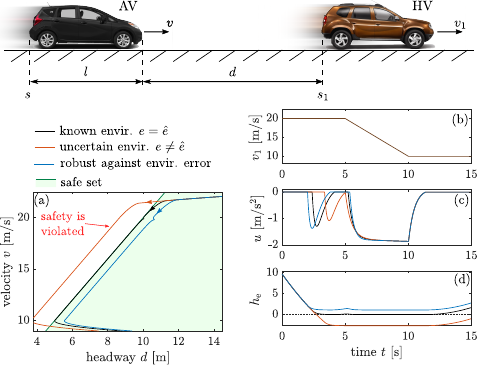}}
\caption{
Strategies for adaptive cruise control where an automated vehicle (AV) intends to safely follow a human-driven vehicle (HV).
The HV represents an environment for the AV.
In the ideal scenario where the HV's position and speed are accurately known to the AV, controller~(\ref{eq:QP_environment}) ensures safety (black).
When the HV's position and speed are measured with error (the environment is uncertain), controller~(\ref{eq:QP_environment}) violates safety (red).
When controller~(\ref{eq:QP_environment}) is robustified via constraint~(\ref{eq:safety_condition_mrcbf_environment}), safety is maintained even in the presence of environment uncertainty (blue).
}
\label{fig:ACC_nodelay}
\end{figure}

\begin{exmp}[Adaptive Cruise Control] \label{ex:ACC_nodelay}
We consider an adaptive cruise control (ACC) problem, where an automated vehicle (AV) intends to follow a human-driven vehicle (HV) without collision; see Fig.~\ref{fig:ACC_nodelay}.
This problem has been studied previously without the notion of ECBFs.
In~\cite{Niletal2016} polyhedral controlled invariant sets and finite-state abstraction were used, and in~\cite{AmesXuGriTab2017, HeOrosz2018} CBFs were applied.
We revisit this problem, use ECBFs to tackle it, and demonstrate that this framework allows us to explicitly take into account uncertainties in the HV's motion.
This will also play an essential role in Section~\ref{sec:delay} to extend the resulting safety-critical controller to safe ACC with input delay, which was not addressed in~\cite{Niletal2016, AmesXuGriTab2017, HeOrosz2018}.

We denote the length of the AV by $l$, the position of its rear bumper by $s$ and its speed by $v$, and we model its motion by:
\begin{equation}
\underbrace{\begin{bmatrix}
\dot{s} \\
\dot{v} \\
\end{bmatrix} }_{\dot{x}}
= 
\underbrace{
\begin{bmatrix}
v \\
-p(v) \\
\end{bmatrix}}_{f(x)}
+ 
\underbrace{
\begin{bmatrix}
0 \\
1 \\
\end{bmatrix}}_{g(x)} u,
\label{eq:ACC_nodelay}
\end{equation}
where ${p(v)}$
indicates resistance terms.
The input $u$ is acceleration command that is assumed to be realized by a low-level controller.
The HV's position and speed, denoted by $s_{1}$ and $v_{1}$,
characterize the environment for the AV: ${e = s_{1}}$ and ${\dot{e} = v_{1}}$.

To avoid collisions, the AV intends to keep its speed $v$ below a safe limit ${\bar{\kappa} d}$ for a selected ${\bar{\kappa} > 0}$, where this limit depends on the distance ${d = s_{1} - s - l}$.
Thus, we use the ECBF:
\begin{equation}
H(x,e) = \bar{\kappa}(s_{1} - s - l) - v,
\end{equation}
and a linear extended class $\mathcal{K}$ function ${\alpha(h) = \gamma h}$ with ${\gamma>0}$.
For this choice, we have
${\dHdx(x,e) f(x) = -\bar{\kappa} v + p(v)}$,
${\dHdx(x,e) g(x) = -1}$ and
${\dHde(x,e) \dot{e} = \bar{\kappa} v_{1}}$.

Substituting these expressions into~(\ref{eq:hdot_environment}) and the safety condition~(\ref{eq:safety_condition_environment}) leads to:
\begin{equation}
\bar{\kappa}(v_{1} - v) + \gamma(\bar{\kappa}(s_{1} - s - l) - v) + p(v) \geq u.
\label{eq:ACC_safety_nodelay}
\end{equation}
Hence the AV should not accelerate more than the expression on the left-hand side.
This expression resembles the desired acceleration of simple ACC controllers, in fact, for ${p(v)=0}$ it is equivalent to the one in~\cite{HeOrosz2018} with a special choice of feedback gains and range policy.
Enforcing~(\ref{eq:ACC_safety_nodelay}), for example, through the QP~(\ref{eq:QP_environment}), guarantees safety based on Theorem~\ref{thm:safety_environment}.

Fig.~\ref{fig:ACC_nodelay} shows numerical simulation results with the safety-critical controller for  ${p(v)=0.1 + 0.0003 v^2}$, ${\gamma=3}$ and ${\bar{\kappa}=2}$ (with units in SI).
The HV performs constant speed cruising, braking with $2\,{\rm m/s^2}$ and constant speed cruising again; see panel (b).
The AV intends to travel at a constant speed higher than the HV's speed with desired controller ${K_{\rm d}(x,e,\dot{e}) = 0}$.
By applying the QP~(\ref{eq:QP_environment}) with the constraint~(\ref{eq:ACC_safety_nodelay}), the AV is able to slow down safely behind the HV; see the black curve.

The controller relies on the position and speed of the HV.
These can be obtained by on-board sensors like radar, lidar, cameras or ultrasonics, or by vehicle-to-vehicle connectivity with the HV.
If these quantities are measured with error, safety may be violated.
This is demonstrated by red color in Fig.~\ref{fig:ACC_nodelay}, where the controller relies on the measured values ${\hat{e} = \hat{s}_{1} = s_{1} + 1\,{\rm m}}$ and ${\hat{\dot{e}} = \hat{v}_{1} = v_{1} + 1\,{\rm m/s}}$ instead of the true values $e=s_{1}$ and $\dot{e} = v_{1}$.
Overestimating the position and speed of the HV causes the system to leave the safe set.

The controller can be made robust to such uncertainties in the environment by replacing the safety condition~(\ref{eq:safety_condition_environment}) with the robustified constraint~(\ref{eq:safety_condition_mrcbf_environment}) in the QP~(\ref{eq:QP_environment}).
If the HV's position and speed estimates have known error bounds $\varepsilon_{s}$ and $\varepsilon_{v}$, that is, ${| s_{1} - \hat{s}_{1} | \leq \varepsilon_{s}}$ and  ${| v_{1} - \hat{v}_{1} | \leq \varepsilon_{v}}$, then, after substitution into~(\ref{eq:safety_condition_mrcbf_environment}) and using~(\ref{eq:robustify_environment}), the robustified constraint becomes:
\begin{equation}
\bar{\kappa}(v_{1} - \varepsilon_{v} - v) + \gamma(\bar{\kappa}(s_{1} - \varepsilon_{s} - s - l) - v) + p(v) \geq u,
\label{eq:ACC_safety_nodelay_robust}
\end{equation}
where we used the Lipschitz coefficients
$\mathcal{L}_{\nabla H f,e} = \mathcal{L}_{\nabla H g,e} = \mathcal{L}_{\nabla H \dot{e},e} = 0$,
${\mathcal{L}_{\alpha \circ H,e} = \gamma \bar{\kappa}}$,
${\mathcal{L}_{\nabla H \dot{e},\dot{e}} = \bar{\kappa}}$.
In this example, the additional robustifying terms are equivalent to considering the worst-case (smallest possible) position and speed for the HV.

The effect of these robustifying terms is shown by blue color in Fig.~\ref{fig:ACC_nodelay} for ${\varepsilon_{s} = 1.4\,{\rm m}}$ and ${\varepsilon_{v} = 1.4\,{\rm m/s}}$.
The AV is able to safely slow down behind the HV despite the uncertainty in the HV's measured state.
Notice that the controller is slightly conservative: the AV stays farther from the boundary of the safe set than in the case without uncertainty.
\end{exmp}

\section{Safety of Systems with Input Delay}
\label{sec:delay}

Now consider the system with input delay ${\tau > 0}$:
\begin{equation}
\dot{x}(t)=f(x(t)) + g(x(t))u(t-\tau),
\label{eq:system_delay}
\end{equation}
where $f$ and $g$ are the same as in~(\ref{eq:system}), and $u$ is bounded and continuous almost everywhere (with a potential discontinuity at ${t = 0}$ when the controller is turned on).
We still assume that there exists a unique solution $x(t)$ over ${t \geq 0}$.

\subsection{Solution of the System and Predictors}

To synthesize safety-critical controllers, we ensure that given the state $x(t)$ at time $t$ the solution of~(\ref{eq:system_delay}) continues to be safe over ${[t,t+\tau]}$.
This property depends on the instantaneous control input $u(t)$ to be synthesized via CBFs and also on the input history over ${[t-\tau,t)}$ given by ${u_t \in \mathcal{B}}$: 
\begin{equation}
u_t(\theta) = u(t+\theta), \quad \theta \in [-\tau,0).
\end{equation}
Here $\mathcal{B}$ denotes the space of functions mapping from $[-\tau,0)$ to $U$ that are bounded and continuous almost everywhere.

The solution over ${[t,t+\tau]}$ is characterized by the {\em semi-flow} ${\Psi: [0,\tau] \times X \times \mathcal{B} \to X}$ as a function of the state $x(t)$ and as a functional of the input history $u_t$:
\begin{equation}
x(t+\vartheta) = \Psi(\vartheta,x(t),u_t), \quad \vartheta \in [0,\tau].
\label{eq:solution}
\end{equation}
The semi-flow is obtained by the forward integration of~(\ref{eq:system_delay}):
\begin{multline}
\Psi(\vartheta,x,u_t) \\
= x \!+\! \int_{0}^{\vartheta} \!\!\!\Big(
f \big( \Psi(\varphi,x,u_t) \big) \!+\!
g \big( \Psi(\varphi,x,u_t) \big) u_t(\varphi\!-\!\tau) \!\Big) {\rm d}\varphi.
\label{eq:semiflow}
\end{multline}
Of particular interest will be
the state $x(t+\tau)$,
that reads:
\begin{equation}
x(t+\tau) = \Psi(\tau,x(t),u_t).
\label{eq:predictedstate}
\end{equation}
We remark that, since $u$ is bounded, $u(t)$ does not affect the value of the integral and thus $u_t$ is defined over $[-\tau,0)$.
That is, the input history $u_t$ does not include the instantaneous control input $u(t)$.
This will allow us to utilize the input history $u_t$ when synthesizing the control input $u(t)$.

Hereinafter, ${x(t+\tau)=\Psi(\tau,x(t),u_t)}$ is called {\em predicted state} and~(\ref{eq:semiflow}) serves as {\em predictor}.
The predicted state will play a key role in safety-critical control.
It can be calculated by forward integration of~(\ref{eq:system_delay}) over ${[t,t+\tau]}$.
Explicit expressions are available for linear systems with ${A \in \mathbb{R}^{n \times n}}$,  ${B \in \mathbb{R}^{n \times m}}$:
\begin{equation}
\dot{x}(t)=A x(t) + B u(t-\tau),
\end{equation}
where the predicted state is given by the convolution integral:
\begin{equation}
\Psi(\tau,x(t),u_t) = {\rm e}^{A\tau}x(t) + \int_{0}^{\tau} \!\! {\rm e}^{A(\tau-\vartheta)} B u_t(\vartheta-\tau){\rm d}\vartheta.
\end{equation}
Note that predictors also exist for systems with time-varying and state-dependent delay, as given in~\cite{BekKrs2013}, that have also been considered in the context of safety in~\cite{Abel2021}.
While the upcoming theorems are stated for constant delay, they could be extended to varying delays by using the appropriate predictor.

\subsection{Control Barrier Functions with Input Delay}

The following definition generalizes CBFs for systems with input delay in the form~(\ref{eq:system_delay}) with ${\tau > 0}$.
\begin{defn}\label{defn:CBF_delay}
Function $h$ is a control barrier function (CBF) for~(\ref{eq:system_delay}) with ${\tau>0}$ if there exists an extended class $\mathcal{K}$ function $\alpha$ such that for all ${x \in S}$ and ${u_t \in \mathcal{B}}$:
\begin{equation}
\sup_{u \in U} \dot{h}(x_{\rm p},u) > - \alpha(h(x_{\rm p})),
\label{eq:CBF_condition_delay}
\end{equation}
where ${x_{\rm p}=\Psi(\tau,x,u_t)}$ with $\Psi$ given by~(\ref{eq:semiflow}).
\end{defn}

The definition recovers Definition~\ref{defn:CBF} in the delay-free case, since ${x_{\rm p}=x}$ if ${\tau=0}$.
With this definition we guarantee safety analogously to Theorem~\ref{thm:safety}.
We assume that safety-critical control starts at ${t=0}$.
According to~(\ref{eq:solution}), ${x(\vartheta) = \Psi(\vartheta,x_{0},u_{0})}$, ${\vartheta \in [0,\tau]}$, that is, the solution over ${[0,\tau]}$ evolves based on the initial input history $u_0$ which we cannot prescribe.
Therefore, we need the following assumption to ensure safety over ${[0,\tau]}$.
\begin{assump}\label{assump:history}
The initial history $u_0$ of the control input satisfies ${x(\vartheta)=\Psi(\vartheta,x_{0},u_{0}) \in S}$, ${\forall \vartheta \in [0,\tau]}$.
\end{assump}
Now we are ready to state our main theorem that ensures safety in the presence of the input delay ${\tau>0}$.

\begin{thm}\label{thm:safety_delay}
\textit{
If $h$ is a CBF for~(\ref{eq:system_delay}) with ${\tau>0}$, then any locally Lipschitz continuous controller ${u=k(x_{\rm p})}$, ${x_{\rm p}=\Psi(\tau,x,u_t)}$ with input history $u_t$ satisfying:
\begin{equation}
\dot{h}(x_{\rm p},u) \geq - \alpha(h(x_{\rm p})),
\label{eq:safety_condition_delay}
\end{equation}
${\forall x \in S}$ and ${\forall u_t \in \mathcal{B}}$ renders $S$ forward invariant under Assumption~\ref{assump:history}, i.e., it ensures ${x_{0} \in S \Rightarrow x(t) \in S}$, ${\forall t \geq 0}$.
}
\end{thm}

\proof
Since Assumption~\ref{assump:history} ensures ${x(\vartheta) \in S}$, ${\forall \vartheta \in [0,\tau]}$, it is sufficient to prove ${x(\tau) \in S \Rightarrow x(t) \in S}$, ${\forall t \geq \tau}$.
By differentiation of~(\ref{eq:semiflow}) with respect to $\vartheta$ we have:
\begin{multline}
\frac{{\rm d}}{{\rm d}\vartheta} \Psi(\vartheta,x(t),u_t) = f \big( \Psi(\vartheta,x(t),u_t) \big) \\
+ g \big( \Psi(\vartheta,x(t),u_t) \big) u(t + \vartheta - \tau).
\label{eq:solution_derivative}
\end{multline}
Furthermore, by noticing ${\frac{{\rm d}}{{\rm d}\vartheta}x(t+\vartheta) = \frac{{\rm d}}{{\rm d}t}x(t+\vartheta)}$ and by using~(\ref{eq:solution}), we get ${\frac{{\rm d}}{{\rm d}\vartheta}\Psi(\vartheta,x(t),u_t) = \frac{{\rm d}}{{\rm d}t}\Psi(\vartheta,x(t),u_t)}$.
Substituting this into~(\ref{eq:solution_derivative}) and using ${\vartheta=\tau}$, we get the following delay-free system for ${x_{\rm p}(t) = \Psi(\tau,x(t),u_t)}$:
\begin{equation}
\dot{x}_{\rm p}(t)=f(x_{\rm p}(t)) + g(x_{\rm p}(t))u(t).
\label{eq:system_pred}
\end{equation}
For this system, Theorem~\ref{thm:safety} can be applied since~(\ref{eq:CBF_condition_delay},~\ref{eq:safety_condition_delay}) hold, thus we get ${x_{\rm p}(0) \in S \Rightarrow x_{\rm p}(t) \in S}$, ${\forall t \geq 0}$ that is equivalent to ${x(\tau) \in S \Rightarrow x(t) \in S}$, ${\forall t \geq \tau}$.
\hfill \qedsymbol

\begin{rem}
As opposed to the delay-free case, the controller in Theorem~\ref{thm:safety_delay} is no longer a state-feedback controller, but it also depends on the input history $u_t$ through the predicted state ${x_{\rm p}=\Psi(\tau,x,u_t)}$.
Furthermore, optimization-based controllers for systems with input delay can be synthesized via Theorem~\ref{thm:safety_delay} similarly to~(\ref{eq:QP}).
The following QP can be solved if ${\tau>0}$:
\begin{align}
\begin{split}
k(x_{\rm p}) =
\mathrm{arg}\hspace{-.1cm}\min_{\hspace{-.2cm} u \in \mathbb{R}^m} & \quad \|u - k_{\rm d}(x_{\rm p})\|^2  \\
\mathrm{s.t.} & \quad \dot{h}(x_{\rm p},u) \geq - \alpha(h(x_{\rm p})).
\end{split}
\label{eq:QP_delay}
\end{align}
Here the desired controller ${k_{\rm d} : X \to U}$ may also account for the delay and can potentially depend on the predicted state.
The solution to~(\ref{eq:QP_delay}) is equivalent to applying the control law~(\ref{eq:QP}) of the corresponding delay-free system on the predicted state ${x_{\rm p}=\Psi(\tau,x,u_t)}$.
This allows one to extend {\em explicitly} available delay-free control laws, such as~(\ref{eq:QP_solution}), for systems with input delays.
However, an explicit expression for $k$ is not always available, especially if additional constraints are added to~(\ref{eq:QP}).
In such cases, one cannot construct ${u}$ by separately solving the delay-free QP~(\ref{eq:QP}) and calculating the predicted state $x_{\rm p}$, but one needs to solve QP~(\ref{eq:QP_delay}) directly.
\end{rem}

\begin{rem} \label{rem:prediction_error}
In practice, predicting the future state may not be perfectly accurate.
Often only an estimate $\hat{x}_{\rm p}$ of the predicted state $x_{\rm p}$ is available.
Classically, this estimate is provided by the numerical forward integration of~(\ref{eq:system_delay}).
Alternatively, state prediction can also be done by more modern tools such as data-driven methods and machine learning.
Theorem~\ref{thm:safety_delay} guarantees safety for the ideal scenario of accurate prediction, ${\hat{x}_{\rm p}=x_{\rm p}}$.
However, mismatches between ${\hat{x}_{\rm p}}$ and ${x_{\rm p}}$ inevitably occur due to model uncertainties and computation errors~\cite{Karafyllis2017},
and longer prediction (larger delay) typically yields larger {\em prediction error} ${\hat{x}_{\rm p}-x_{\rm p}}$.
The effect of prediction errors can be studied via the notion of input-to-state safety~\cite{ames2019issf, Alan2022},
as we did in~\cite{Molnar2021lcss}, where we showed that  the input disturbance ${d = k(\hat{x}_{\rm p})-k(x_{\rm p})}$ makes a larger set $S_{\rm d} \supseteq S$ forward invariant.
Alternatively, if the prediction error is bounded and there exists ${\varepsilon_x > 0}$ such that
${\| \hat{x}_{\rm p} - x_{\rm p} \| \leq \varepsilon_{x}}$,
robustness against the prediction error can be provided analogously to Proposition~\ref{prop:mrcbf_environment}
using the following condition:
\begin{equation}
\dot{h}(\hat{x}_{\rm p},u)
-(\mathcal{L}_{\nabla h f} + \mathcal{L}_{\alpha \circ h}) \varepsilon_{x} - \mathcal{L}_{\nabla h g} \varepsilon_{x} \| u \| \geq - \alpha \big( h (\hat{x}_{\rm p}) \big),
\end{equation}
where $\mathcal{L}$ is the Lipschitz coefficient of the subscripted function on $S$.
This method was originally used in~\cite{Dean2021} to address mismatches between estimated and true states ($\hat{x}$ and $x$) of delay-free systems.
While safety is guaranteed, the additional terms may lead to conservative behavior where the system evolves far away from the safe set boundary.
The conservatism depends on the error bound $\varepsilon_x$ and the Lipschitz coefficients.
\end{rem}

\subsection{Safety with Input Delay in Dynamic Environment}

Finally, we consider the scenario when safety needs to be guaranteed for the time delay system~(\ref{eq:system_delay}) in a dynamic environment described by the state $e(t)$ and the environmental safe set $S_{\rm e}$.
We assume that the state $e(t)$ of the environment is a continuously differentiable function of time\footnote{In case of a higher relative degree ${r>1}$, the state $e(t)$ of the environment must be $r$ times continuously differentiable. While we omit in-depth discussion about higher relative degrees, an example is shown in Section~\ref{sec:segway}.}.

In Theorem~\ref{thm:safety_delay}, the key step to achieve safety was to predict the system's state over the time interval $[t,t+\tau]$.
Now we make a {\em prediction of the environment} and rely on its future state ${e_{\rm p}(t) = e(t+\tau)}$.
Typically, the future state $e_{\rm p}(t)$ depends on the current state $e(t)$, as emphasized by the notation:
\begin{equation}
e(t+\vartheta) = \Gamma(\vartheta,e(t)), \quad \vartheta \in [0,\tau],
\label{eq:map_environment}
\end{equation}
where the map ${\Gamma: [0,\tau] \times E \to E}$ may be unknown and may involve dependence on other quantities as well.

For this setup, we establish safety by extending the notion of ECBFs to systems with input delay.

\begin{defn}\label{defn:CBF_delay_environment}
Function $H$ is an environmental control barrier function (ECBF) for~(\ref{eq:system_delay}) with ${\tau>0}$ if there exists an extended class $\mathcal{K}$ function $\alpha$ such that for all ${(x,e) \in S_{\rm e}}$, ${\dot{e} \in \mathcal{E}}$ and ${u_t \in \mathcal{B}}$:
\begin{equation}
\sup_{u \in U} \dot{H}(x_{\rm p},e_{\rm p},\dot{e}_{\rm p},u) > - \alpha(H(x_{\rm p},e_{\rm p})),
\label{eq:CBF_condition_delay_environment}
\end{equation}
where ${x_{\rm p}=\Psi(\tau,x,u_t)}$ with $\Psi$ given by~(\ref{eq:semiflow}),
while ${e_{\rm p} = \Gamma(\tau,e)}$ with $\Gamma$ defined by~(\ref{eq:map_environment}).
\end{defn}

With this definition, Theorems~\ref{thm:safety_environment} and~\ref{thm:safety_delay}, that separately guarantee safety in dynamic environment and for input delay, can be integrated into Theorem~\ref{thm:safety_delay_environment} below.
Again, we make a preliminary assumption that the system is safe over the interval ${t \in [0,\tau]}$ when safety depends on the initial input history $u_{0}$.

\begin{assump}\label{assump:history_environment}
The initial history $u_0$ of the control input satisfies ${(x(\vartheta),e(\vartheta))=(\Psi(\vartheta,x_{0},u_{0}),\Gamma(\vartheta,e_{0})) \in S_{\rm e}}$, ${\forall \vartheta \in [0,\tau]}$.
\end{assump}

Now we can state the main theorem to ensure safety for systems with input delay in dynamic environment.

\begin{thm}\label{thm:safety_delay_environment}
\textit{
If $H$ is an ECBF for~(\ref{eq:system_delay}) with ${\tau>0}$, then any locally Lipschitz continuous controller
${u=K(x_{\rm p},e_{\rm p},\dot{e}_{\rm p})}$,
${x_{\rm p}=\Psi(\tau,x,u_t)}$,
${e_{\rm p} = \Gamma(\tau,e)}$
with history $u_t$ satisfying:
\begin{equation}
\dot{H}(x_{\rm p},e_{\rm p},\dot{e}_{\rm p},u) \geq - \alpha(H(x_{\rm p},e_{\rm p})),
\label{eq:safety_condition_delay_environment}
\end{equation}
${\forall (x,e) \in S_{\rm e}}$, ${\forall \dot{e} \in \mathcal{E}}$ and ${\forall u_t \in \mathcal{B}}$ renders $S_{\rm e}$ forward invariant under Assumption~\ref{assump:history_environment}, i.e., it ensures ${(x_{0},e_{0}) \in S_{\rm e} \Rightarrow (x(t),e(t)) \in S_{\rm e}}$, ${\forall t \geq 0}$.
}
\end{thm}

\proof
Assumption~\ref{assump:history_environment} yields $(x_{0},e_{0}) \in S_{\rm e} \Rightarrow (x(\vartheta),e(\vartheta)) \in S_{\rm e}$, ${\forall \vartheta \in [0,\tau]}$, thus what remains to prove is ${(x(\tau),e(\tau)) \in S_{\rm e} \Rightarrow (x(t),e(t)) \in S_{\rm e}}$, ${\forall t \geq \tau}$.
This is equivalent to ${(x_{\rm p}(0),e_{\rm p}(0)) \in S_{\rm e} \Rightarrow (x_{\rm p}(t),e_{\rm p}(t)) \in S_{\rm e}}$, ${\forall t \geq 0}$ based on the definitions of $x_{\rm p}$ and $e_{\rm p}$.
According to the proof of Theorem~\ref{thm:safety_delay}, $x_{\rm p}(t)$ is governed by the delay-free dynamics~(\ref{eq:system_pred}).
Hence, Theorem~\ref{thm:safety_environment} is directly applicable to this delay-free system considering the environment given by $e_{\rm p}$.
This provides ${(x_{\rm p}(0),e_{\rm p}(0)) \in S_{\rm e} \Rightarrow (x_{\rm p}(t),e_{\rm p}(t)) \in S_{\rm e}}$, ${\forall t \geq 0}$ as desired, which completes the proof.
\hfill \qedsymbol

\begin{rem}
Theorem~\ref{thm:safety_delay_environment} ultimately leads to controllers that use the state $x$, the input history $u_t$, and the state of the environment given by $e$, $\dot{e}$.
An example is the following QP:
\begin{align}
\begin{split}
K(x_{\rm p},e_{\rm p},\dot{e}_{\rm p}) =
\mathrm{arg}\hspace{-.1cm}\min_{\hspace{-.2cm} u \in U} & \quad \|u - K_{\rm d}(x_{\rm p},e_{\rm p},\dot{e}_{\rm p})\|^2  \\
\mathrm{s.t.} & \quad \dot{H}(x_{\rm p},e_{\rm p},\dot{e}_{\rm p},u) \geq - \alpha(H(x_{\rm p},e_{\rm p})),
\end{split}
\label{eq:QP_delay_environment}
\end{align}
with
${x_{\rm p}=\Psi(\tau,x,u_t)}$ and
${e_{\rm p} = \Gamma(\tau,e)}$,
cf.~(\ref{eq:QP_environment},~\ref{eq:QP_delay}).
\end{rem}

\begin{rem}
In practice, the environment's future state $e_{\rm p}$ and its derivative $\dot{e}_{\rm p}$ are unknown, and we can only provide estimates $\hat{e}_{\rm p}$ and $\hat{\dot{e}}_{\rm p}$.
Robustness against {\em environment prediction errors} is a significant problem since the evolution of the environment is typically more uncertain than the dynamics of the control system.
Robustness can be addressed similarly to Section~\ref{sec:ECBF_robustness}, as follows.
For simplicity, we assume that the dynamics of the control system~(\ref{eq:system_delay}) is well-known and its state is predicted with negligible error (${\hat{x}_{\rm p}=x_{\rm p}}$); otherwise prediction errors could be overcome based on Remark~\ref{rem:prediction_error}.
Then, the approach of Proposition~\ref{prop:mrcbf_environment} can be applied to achieve robustness against environment prediction errors, via the condition:
\begin{equation}
\dot{H}(x_{\rm p},\hat{e}_{\rm p},\hat{\dot{e}}_{\rm p},u)
- C(\varepsilon_{e},\varepsilon_{\dot{e}},u) \geq - \alpha \big( H (x_{\rm p},\hat{e}_{\rm p}) \big),
\label{eq:safety_condition_mrcbf_delay_environment}
\end{equation}
with $C(\varepsilon_{e},\varepsilon_{\dot{e}},u)$ defined in~(\ref{eq:robustify_environment}), where $\varepsilon_{e}$ and $\varepsilon_{\dot{e}}$ are error bounds satisfying ${\big\| e_{\rm p} - \hat{e}_{\rm p} \big\| \leq \varepsilon_{e}}$ and ${\big\| \dot{e}_{\rm p} - \hat{\dot{e}}_{\rm p} \big\| \leq \varepsilon_{\dot{e}}}$.
\end{rem}

\begin{exmp}[Adaptive Cruise Control with input delay] \label{ex:ACC_delay}
Consider the adaptive cruise control problem of Example~\ref{ex:ACC_nodelay}, now with input delay $\tau$ that represents powertrain delays:
\begin{equation}
\underbrace{\begin{bmatrix}
\dot{s}(t) \\
\dot{v}(t) \\
\end{bmatrix} }_{\dot{x}(t)}
= 
\underbrace{
\begin{bmatrix}
v(t) \\
-p\big(v(t)\big) \\
\end{bmatrix}}_{f(x(t))}
+ 
\underbrace{
\begin{bmatrix}
0 \\
1 \\
\end{bmatrix}}_{g(x(t))} u(t-\tau).
\label{eq:ACC_delay}
\end{equation}
For passenger vehicles, the delay $\tau$ is around 0.5--1 s~\cite{Ji2021}, hence it is not negligible for safety-critical applications.

\begin{figure}[!t]
\centerline{\includegraphics[width=\columnwidth]{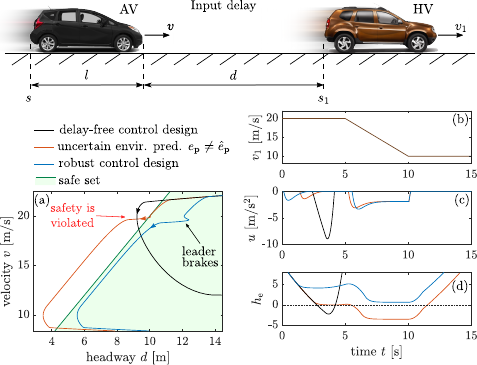}}
\caption{
Strategies for adaptive cruise control with input delay.
The na\"ive implementation of the delay-free control design~(\ref{eq:QP_environment}) violates safety (black).
The controller~(\ref{eq:QP_delay_environment}) that relies on predictor feedback enforces safety as long as the prediction of the HV's motion is accurate, and violates safety otherwise (red).
When controller~(\ref{eq:QP_delay_environment}) is robustified via constraint~(\ref{eq:safety_condition_mrcbf_delay_environment}), safety is maintained despite prediction errors (blue).
}
\label{fig:ACC_delay}
\end{figure}

The effect of the delay is demonstrated in Fig.~\ref{fig:ACC_delay} by black color.
Simulation results are shown with a large delay ${\tau = 1}$, zero initial input history, and the parameters of Example~\ref{ex:ACC_delay}:
${p(v)=0.1 + 0.0003 v^2}$,
${\gamma=3}$ and ${\bar{\kappa}=2}$ (with units in SI).
If one implements the delay-free control design~(\ref{eq:QP_environment}) relying on~(\ref{eq:ACC_safety_nodelay}), it fails to keep system~(\ref{eq:ACC_delay}) safe due to the delay $\tau$.
Safety is violated even when the HV cruises at constant speed.

Thus, we use Theorem~\ref{thm:safety_delay_environment} to ensure safety for ${\tau>0}$.
We predict the AV's motion by forward integrating~(\ref{eq:ACC_delay}) over the delay interval ${[t,t+\tau]}$ using the input history $u_t$.
The resulting predicted state is denoted by ${x_{\rm p}(t) = [s_{\rm p}(t),\,v_{\rm p}(t)]^\top}$.
Furthermore, we predict the HV's motion by assuming constant speed over ${[t,t+\tau]}$: ${\hat{\dot{e}}_{\rm p}(t) = v_{1}(t)}$ and ${\hat{e}_{\rm p}(t) = s_{1}(t) + v_{1}(t) \tau}$.
The prediction is incorporated into the safety condition~(\ref{eq:safety_condition_delay_environment}) to synthesize a control input satisfying:
\begin{equation}
\bar{\kappa} \big( v_{1} - v_{\rm p} \big) + \gamma \big( \bar{\kappa} ( s_{1} + v_{1} \tau - s_{\rm p} - l ) - v_{\rm p} \big) + p(v_{\rm p}) \geq u,
\label{eq:ACC_safety_delay}
\end{equation}
cf.~(\ref{eq:ACC_safety_nodelay}).

Red color in Fig.~\ref{fig:ACC_delay} shows the result of executing the corresponding controller given by QP~(\ref{eq:QP_delay_environment}) with desired controller ${K_{\rm d}(x_{\rm p},e_{\rm p},\dot{e}_{\rm p}) = 0}$ and constraint~(\ref{eq:ACC_safety_delay}).
The controller maintains safety as long as the HV travels at constant speed and the prediction about HV's future motion is accurate (${\hat{\dot{e}}_{\rm p}(t) = \dot{e}_{\rm p}(t)}$ and ${\hat{\dot{e}}_{\rm p}(t) = \dot{e}_{\rm p}(t)}$).
Then, safety is violated once the HV starts to slow down and the prediction no longer matches the true future motion of the HV (${\hat{\dot{e}}_{\rm p}(t) \neq \dot{e}_{\rm p}(t)}$ and ${\hat{\dot{e}}_{\rm p}(t) \neq \dot{e}_{\rm p}(t)}$).
While one can argue that the constant speed prediction is overly simplistic and more sophisticated predictions exist, the HV's future motion is inherently uncertain.
Hence, we need to robustify the controller against this uncertainty.

Predicting the HV's future motion with constant speed leads to a time-varying prediction error that depends on the velocity profile $v_{1}(t)$.
Assuming that the HV's acceleration is limited to a range ${[-a_{\rm min}, a_{\rm max}]}$, we have the following physical bounds for the environment prediction error: ${\varepsilon_{\rm \dot{e}} = \bar{a} \tau}$ and ${\varepsilon_{\rm e} = \bar{a} \tau^2/2}$ with ${\bar{a} = \max \{ a_{\rm min}, a_{\rm max} \}}$.
Then the robustified condition~(\ref{eq:safety_condition_mrcbf_delay_environment}) leads to the form:
\begin{multline}
\bar{\kappa} \big( v_{1} - \bar{a} \tau - v_{\rm p} \big) + \gamma \big( \bar{\kappa} ( s_{1}  + v_{1} \tau - \bar{a} \tau^2 / 2 - s_{\rm p} - l ) - v_{\rm p} \big) \\ + p(v_{\rm p}) \geq u,
\label{eq:ACC_safety_delay_robust}
\end{multline}
cf.~(\ref{eq:ACC_safety_delay}).
Besides, it can be shown that replacing $\bar{a}$ with $a_{\rm min}$ in~(\ref{eq:ACC_safety_delay_robust}) also implies~(\ref{eq:safety_condition_delay_environment}).
This provides a problem-specific bound that is less conservative than~(\ref{eq:ACC_safety_delay_robust}) if ${a_{\rm min}<a_{\rm max}}$.

The blue curve in Fig.~\ref{fig:ACC_delay} shows simulation results for controller~(\ref{eq:QP_delay_environment}) with the robustified constraint~(\ref{eq:ACC_safety_delay_robust}) and $a_{\rm min} = a_{\rm max} = 2.5\,{\rm m/s^2}$.
By Theorem~\ref{thm:safety_environment}, the controller ensures safety even with input delay, in a dynamic, uncertain environment.
The price of robustness is slight conservatism: the system does not reach the safe set boundary but keeps a small distance, since the controller uses the $2.5\,{\rm m/s^2}$ braking limit instead of the actual $2\,{\rm m/s^2}$ braking over the ${1\,{\rm s}}$ delay interval.
\end{exmp}

\section{Case-Study: Control of a Segway}
\label{sec:segway}

\begin{figure}[!t]
\centerline{\includegraphics[scale=1]{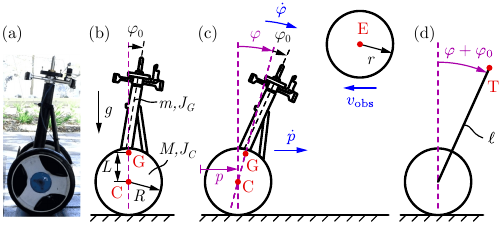}}
\caption{
(a) The Ninebot E+ Segway platform.
(b) Its mechanical model illustrated in equilibrium position.
(c) The Segway in motion aiming to avoid a moving obstacle.
(d) Simplified representation of the Segway.
}
\label{fig:segway}
\end{figure}

Now we apply the theoretical constructions of this paper to a real-life robotic system: we consider the control of a Ninebot E+ Segway platform~\cite{Gurriet2018a}
shown in Fig.~\ref{fig:segway}(a).
We intend to drive the Segway so that it safely avoids a moving obstacle, even when the obstacle position is uncertain and there is a delay in the control loop.
We conduct numerical simulations of the Segway's motion using a high-fidelity dynamical model.

We describe the planar motion of the Segway by
its mechanical model in Fig.~\ref{fig:segway}(b,c).
Fig.~\ref{fig:segway}(b) depicts the Segway in equilibrium, where the center of mass of its frame (point G) is above the wheel center (point C).
Note that the frame is asymmetric and its axis is tilted in equilibrium at an offset angle $\varphi_{0}$.
Fig.~\ref{fig:segway}(c) shows the Segway in motion during obstacle avoidance, and Fig.~\ref{fig:segway}(d) depicts its simplified representation.

Our goal is to drive the Segway forward with a desired speed $\dot{p}_{\rm d}$ while avoiding a moving, circular obstacle centered at ${[e,y]^\top}$ (point E in Fig.~\ref{fig:segway}) with radius $r$.
The obstacle represents the environment of the Segway.
We intend to control the Segway such that its tip -- point T in Fig.~\ref{fig:segway}, located at distance $\ell$ from the wheel center -- does not collide with the obstacle.
The obstacle moves horizontally with constant speed $v_{\rm obs}$:
${e = e_{0} -  t v_{\rm obs}}$, ${\dot{e} = -v_{\rm obs}}$, ${\dot{y} = 0}$.
For numerical case-study, we use
${\dot{p}_{\rm d}=1\,{\rm m/s}}$,
${r=0.2\,{\rm m}}$,
${v_{\rm obs}=0.5\,{\rm m/s}}$,
${e_{0}=1\,{\rm m}}$
and
${y=1.0418\,{\rm m}}$ (for this value, point T is located $0.05\,{\rm m}$ above the bottom of the obstacle when the Segway is in equilibrium).
First, we consider safety-critical control by neglecting the time delay that may arise in the Segway's control loop, then we address the effects of delay.

\subsection{Safety-Critical Control in Dynamic Environment}

\begin{figure}
\centerline{\includegraphics[width=\columnwidth]{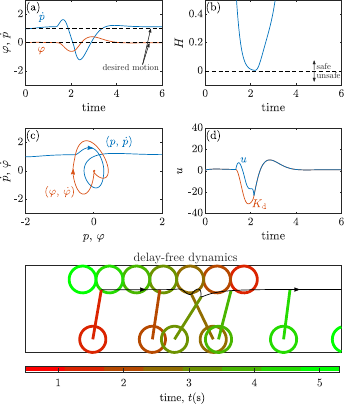}}
\caption{
Safety-critical control of the Segway to avoid a moving obstacle, with
the delay-free dynamics~(\ref{eq:segway_dynamics}) and known obstacle position.
The Segway safely avoids the obstacle with a controller that satisfies~(\ref{eq:segway_safety_condition}).
}
\label{fig:segway_reference}
\end{figure}

We describe the Segway dynamics by the wheel center position $p$ and pitch angle $\varphi$ as two-degrees-of-freedom planar system with general coordinates ${q = [p,\,\varphi]^\top \in \mathcal{Q}}$ and velocities ${\dot{q} = [v,\,\omega]^\top \in \mathbb{R}^{2}}$.
The state becomes ${x =[p,\,\varphi,\,v,\,\omega]^\top \in X}$, the configuration space is ${\mathcal{Q} = \mathbb{R} \times [0,2\pi]}$, and the state space is ${X = \mathcal{Q} \times \mathbb{R}^{2}}$.
The control input ${u \in \mathbb{R}}$ is the voltage applied on the motors at the wheels.
The dynamics are governed by:
\begin{equation}
\begin{bmatrix}
\dot{p} \\ \dot{\varphi} \\ \dot{v} \\ \dot{\omega}
\end{bmatrix} =
\begin{bmatrix}
v \\ \omega \\ f_{v}(\varphi,v,\omega) \\
f_{\omega}(\varphi,v,\omega)
\end{bmatrix} +
\begin{bmatrix}
0 \\ 0 \\ g_{v}(\varphi) \\ g_{\omega}(\varphi)
\end{bmatrix} u.
\label{eq:segway_dynamics}
\end{equation}
For the derivation of this equation and the detailed expressions of $f_{v}$, $f_{\omega}$, $g_{v}$ and $g_{\omega}$, please refer to Appendix~\ref{sec:appdx_segway_dynamics}.
The model parameters were identified in~\cite{Gurriet2018a}
and are listed in Table~\ref{tab:segway}.

We track the desired speed $\dot{p}_{\rm d}$ by the desired controller:
\begin{equation}
K_{\rm d}(x,e,\dot{e}) = K_{\dot{p}} (\dot{p} - \dot{p}_{\rm d}) + K_{\varphi} \varphi + K_{\dot{\varphi}} \dot{\varphi}
\label{eq:segway_desired_controller}
\end{equation}
with gains
${K_{\dot{p}} = 8\,{\rm Vs/m}}$,
${K_{\varphi} = 40\,{\rm V/rad}}$,
${K_{\dot{\varphi}} = 10\,{\rm Vs/rad}}$,
that also stabilizes the Segway to the upright position.
To avoid the moving obstacle, we construct the ECBF candidate:
\begin{equation}
\begin{split}
H(x,e) & = d^\top d  - r^2, \\
d & =
\begin{bmatrix}
p  + \ell \sin (\varphi+\varphi_{0})  - e \\
R+\ell \cos (\varphi+\varphi_{0})  - y
\end{bmatrix},
\end{split}
\label{eq:segway_ECBF_candidate}
\end{equation}
where $d$ points from the obstacle center to the Segway's tip.


We seek to maintain safety with respect to the environmental safe set~(\ref{eq:safeset_environment}) using Theorem~\ref{thm:safety_environment}.
However, $H$ is not a valid ECBF since ${\dHdx(x,e)  g(x)=0}$ and $\dot{H}$ is independent of the input $u$.
Thus, we use a dynamic extension of the ECBF based on~\cite{Nguyen2016}.
We define the {\em extended environmental control barrier function}:
\begin{equation}
H_{\rm e}(x,e,\dot{e}) = \dot{H}(x,e,\dot{e}) + \gamma_{\rm e} H(x,e),
\label{eq:segway_ECBF_extension}
\end{equation}
with ${\gamma_{\rm e}>0}$, whose derivative depends on the control input $u$:
\begin{multline}
\dot{H}_{\rm e}(x,e,\dot{e},\ddot{e},u) =
\nabla_{x} \dot{H}(x,e,\dot{e}) (f(x)+g(x)u) \\
+ \nabla_{e} \dot{H}(x,e,\dot{e}) \dot{e} 
+ \nabla_{\dot{e}} \dot{H}(x,e,\dot{e}) \ddot{e}
+\gamma_{\rm e} \dot{H}(x,e,\dot{e}).
\end{multline}
With this choice, ${H_{\rm e}(x,e,\dot{e}) \geq 0}$ is equivalent to~(\ref{eq:safety_condition_environment}) in Theorem~\ref{thm:safety_environment} considering a linear class $\mathcal{K}$ function with gradient $\gamma_{\rm e}$.
Thus, safety is achieved if $H_{\rm e}$ is kept nonnegative for all time, which can be enforced if ${H_{\rm e}(x_{0},e_{0},\dot{e}_{0}) \geq 0}$ and:
\begin{equation}
\dot{H}_{\rm e}(x,e,\dot{e},\ddot{e},u) \geq - \alpha \big( H_{\rm e} (x,\hat{e},\dot{e}) \big),
\label{eq:segway_safety_condition}
\end{equation}
cf.~Theorem~\ref{thm:safety_environment}.
Notice that the second derivative $\ddot{e}$ shows up.

We implement a QP-based controller similar to~(\ref{eq:QP_environment}), with desired controller~(\ref{eq:segway_desired_controller}) and constraint~(\ref{eq:segway_safety_condition}) using linear class $\mathcal{K}$ function ${\alpha(H_{\rm e}) = \gamma H_{\rm e}}$ with ${\gamma=7.5\,{\rm s^{-1}}}$ and
${\gamma_{\rm e}=7.5\,{\rm s^{-1}}}$.
For known obstacle position, the performance of the controller is demonstrated in Fig.~\ref{fig:segway_reference}, with snapshots of the motion at the bottom and its characteristics at the top.
Panel (a) shows that the Segway tracks the desired velocity (${\dot{p} \approx \dot{p}_{\rm d}}$) in upright position (${\varphi \approx 0}$) until it has to evade the obstacle.
Panel (b) indicates that the obstacle is safely avoided as $H$ is positive for all time.
Panel (c) shows the corresponding phase portrait, whereas panel (d) depicts the desired and actual control inputs.

\subsection{Safety-Critical Control with Input Delay}

\begin{figure}
\centerline{\includegraphics[width=\columnwidth]{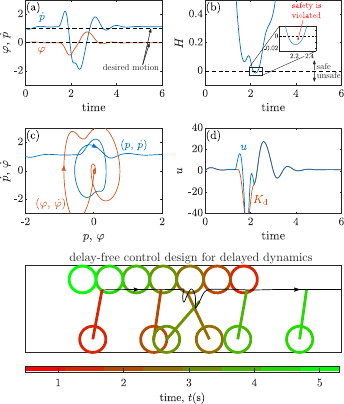}}
\caption{
Safety-critical control of the Segway to avoid a moving obstacle.
The dynamics~(\ref{eq:segway_dynamics_delay}) involve an input delay.
The na\"ive implementation of the delay-free control design based on~(\ref{eq:segway_safety_condition}) fails to avoid the obstacle.
}
\label{fig:segway_delay}
\end{figure}

Now we consider the dynamics with input delay ${\tau>0}$ arising from sensory, feedback and actuation latencies:
\begin{equation}
\begin{bmatrix}
\dot{p}(t) \\ \dot{\varphi}(t) \\ \dot{v}(t) \\ \dot{\omega}(t)
\end{bmatrix} =
\begin{bmatrix}
v(t) \\ \omega(t) \\ f_{v}(\varphi(t),v(t),\omega(t)) \\
f_{\omega}(\varphi(t),v(t),\omega(t))
\end{bmatrix} +
\begin{bmatrix}
0 \\ 0 \\ g_{v}(\varphi(t)) \\ g_{\omega}(\varphi(t))
\end{bmatrix} u(t-\tau),
\label{eq:segway_dynamics_delay}
\end{equation}
cf.~(\ref{eq:segway_dynamics}).
The effect of the delay is illustrated in Fig.~\ref{fig:segway_delay}.
Here the same delay-free control design is used as in Fig.~\ref{fig:segway_reference}, but the dynamics are subject to the input delay ${\tau=0.1\,{\rm s}}$.
Although the Segway realizes a stable motion, the delay leads to safety violation: the Segway collides with the obstacle ($H$ becomes negative in Fig.~\ref{fig:segway_delay}(b)).
While collision could be avoided by buffering the obstacle, formal safety guarantees no longer hold with delay.
Moreover, the control input is much larger with delay than without delay, cf.~Fig.~\ref{fig:segway_reference}(d) and Fig.~\ref{fig:segway_delay}(d).
Such large inputs are undesired as safety-critical control could become infeasible with input bounds.
To overcome the unsafe behavior, the delay needs to be incorporated into the control design.

The input delay can be tackled via predictor feedback, using Theorem~\ref{thm:safety_delay_environment}.
We assume that the state $x_{\rm p}$ is accurately predicted (${\hat{x}_{\rm p}=x_{\rm p}}$), while the predictions ${e_{\rm p}}$, ${\dot{e}_{\rm p}}$ and ${\ddot{e}_{\rm p}}$ of the environment are uncertain.
Hence the controller relies on estimates ${\hat{e}_{\rm p}}$, ${\hat{\dot{e}}_{\rm p}}$ and ${\hat{\ddot{e}}_{\rm p}}$ and their error bounds
${\big\| e_{\rm p} - \hat{e}_{\rm p} \big\| \leq \varepsilon_{e}}$,
${\big\| \dot{e}_{\rm p} - \hat{\dot{e}}_{\rm p} \big\| \leq \varepsilon_{\dot{e}}}$ and
${\big\| \ddot{e}_{\rm p} - \hat{\ddot{e}}_{\rm p} \big\| \leq \varepsilon_{\ddot{e}}}$.
Analogously to~(\ref{eq:safety_condition_mrcbf_delay_environment}), we use the robustified safety constraint:
\begin{equation}
\dot{H}_{\rm e}(x_{\rm p},\hat{e}_{\rm p},\hat{\dot{e}}_{\rm p},\hat{\ddot{e}}_{\rm p},u) - C(\varepsilon_{e},\varepsilon_{\dot{e}},\varepsilon_{\ddot{e}},u) \geq - \alpha \big( H_{\rm e} (x_{\rm p},\hat{e}_{\rm p},\hat{\dot{e}}_{\rm p}) \big),
\label{eq:segway_safety_condition_delay}
\end{equation}
with:
\begin{multline}
C(\varepsilon_{e},\varepsilon_{\dot{e}},u) = (\mathcal{L}_{\nabla H_{\rm e} f,e} + \mathcal{L}_{\alpha \circ H_{\rm e},e} + \mathcal{L}_{\nabla H_{\rm e} \dot{e},e} + \mathcal{L}_{\nabla H_{\rm e} \ddot{e},e}) \varepsilon_{e} \\
+ (\mathcal{L}_{\nabla H_{\rm e} f,\dot{e}} + \mathcal{L}_{\alpha \circ H_{\rm e},\dot{e}} + \mathcal{L}_{\nabla H_{\rm e} \dot{e},\dot{e}} + \mathcal{L}_{\nabla H_{\rm e} \ddot{e},\dot{e}}) \varepsilon_{\dot{e}} \\
+ \mathcal{L}_{\nabla H_{\rm e} \ddot{e},\ddot{e}} \varepsilon_{\ddot{e}} + (\mathcal{L}_{\nabla H_{\rm e} g,e} \varepsilon_{e} + \mathcal{L}_{\nabla H_{\rm e} g,\dot{e}} \varepsilon_{\dot{e}}) \| u \|,
\label{eq:segway_robustify}
\end{multline}
cf.~(\ref{eq:robustify_environment}).
Here $\mathcal{L}$ denotes the Lipschitz coefficient of the subscripted function with respect to the argument at the end of the subscript.
These coefficients were determined based on the expressions of the Segway dynamics; see Appendix~\ref{sec:appdx_segway_Lipschitz}. 

Fig.~\ref{fig:segway_delay_robust} shows the implementation of the corresponding QP-based controller, similar to~(\ref{eq:QP_delay_environment}), with desired controller~(\ref{eq:segway_desired_controller}) applied on the predicted state and with constraint~(\ref{eq:segway_safety_condition_delay}).
The true future of the environment, given by
${\ddot{e}_{\rm p} = 0}$,
${\dot{e}_{\rm p} = - v_{{\rm obs}}}$ and
${e_{\rm p}  = e - v_{{\rm obs}} \tau}$, is unknown to the controller.
Instead, the controller relies on the prediction
${\hat{\ddot{e}}_{\rm p} = 0}$,
${\hat{\dot{e}}_{\rm p} = - (v_{{\rm obs}} - \Delta v)}$ and
${\hat{e}_{\rm p}  = e - (v_{{\rm obs}} - \Delta v) \tau}$.
That is, the speed of the obstacle is underestimated by ${\Delta v = 0.05\,{\rm m/s}}$.
The controller is robustified against the prediction error using the error bounds ${\varepsilon_{\ddot{e}} = 0}$, ${\varepsilon_{\dot{e}} = 0.055\,{\rm m/s}}$ and ${\varepsilon_{e} = \tau \varepsilon_{\dot{e}} = 0.0055\,{\rm m}}$.
With the proposed robust controller, the Segway safely executes the obstacle avoidance task, despite the delay in the control loop and the uncertainty in the obstacle's future position.
This is achieved with a qualitatively different motion than in the delay-free case.
For zero delay in Fig.~\ref{fig:segway_reference}, the Segway pitches backwards to go under the obstacle.
For nonzero delay in Fig.~\ref{fig:segway_delay_robust}, the Segway moves in reverse to get away from the obstacle, then pitches forward to go under it.
Notably, this behavior is automatically generated by ECBF, and with provable guarantees of safety.

\section{Conclusions}
\label{sec:concl}

We have discussed safety-critical control for systems with input delay that operate in dynamically evolving environment.
We have provided formal safety guarantees and proofs thereof.
We have established a method for safe control synthesis by proposing environmental control barrier functions and integrating them with predictor feedback.
We have strengthened the underlying safety condition to provide robustness against uncertain environments, in which the future of the environment cannot be predicted accurately but bounds on the related prediction error are known.
The resulting control design uses worst-case uncertainty bounds and is provably safe.
We have demonstrated the method by an adaptive cruise control problem where the motion of another vehicle creates an uncertain environment, and by a Segway controller that avoids moving obstacles.
Our future work includes the analysis of prediction errors, control of systems with both state and input delays, and use of control barrier functionals acting on delayed states.

\begin{figure}
\centerline{\includegraphics[width=\columnwidth]{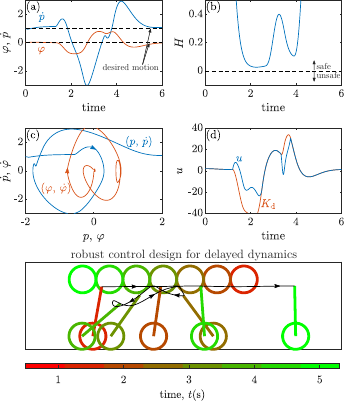}}
\caption{
Safety-critical control of the Segway to avoid a moving obstacle.
The dynamics~(\ref{eq:segway_dynamics_delay}) involve an input delay that is compensated via predictor feedback.
The controller is designed using~(\ref{eq:segway_safety_condition_delay}), taking into account prediction errors.
The Segway successfully avoids the obstacle despite the delay and the uncertain future motion of the obstacle.
}
\label{fig:segway_delay_robust}
\end{figure}


\appendices

\section{KKT Conditions}
\label{sec:appdx_KKT}

This appendix shows the derivation of the solution~(\ref{eq:QP}) to the quadratic program~(\ref{eq:QP_solution}) without input constraints (${U = \mathbb{R}^{m}}$).
Let ${\Delta k(x) = k(x) - k_{\rm d}(x)}$ and consider $\dot{h}(x,u)$ in~(\ref{eq:hdot}) and $\phi_{0}(x)$, $\phi_{1}(x)$ in~(\ref{eq:QP_solution}) with ${\phi_{1}(x) \neq 0}$.
We can restate~(\ref{eq:QP}) as:
\begin{align}
\begin{split}
\Delta k(x) = \mathrm{arg}\hspace{-.15cm}\min_{\hspace{-.2cm} \Delta u \in \mathbb{R}^{m}} & \quad \| \Delta u \|^2  \\
\mathrm{s.t.} & \quad \phi_{0}(x) + \phi_{1}(x) \Delta u \geq 0.
\end{split}
\end{align}

This optimization problem has convex objective and affine constraint, hence the {\em Karush-Kuhn-Tucker (KKT) conditions} \cite{Boyd2004} provide necessary and sufficient conditions for optimality.
The KKT conditions imply that there exists a Lagrange multiplier ${\mu : X \to \mathbb{R}}$ such that $\mu(x)$ and $\Delta k(x)$ satisfy:
\begin{align}
& \mu(x) \geq 0,
\label{eq:kkt_dual_feasibility} \\
& \Delta k(x) = \mu(x) \phi_{1}^\top(x),
\label{eq:kkt_stationary} \\
& \phi_{0}(x) + \phi_{1}(x) \Delta k(x) \geq 0,
\label{eq:kkt_primal_feasibility} \\
& \mu(x) (\phi_{0}(x) + \phi_{1}(x) \Delta k(x)) = 0,
\label{eq:kkt_complementary_slackness}
\end{align}
which are referred to as dual feasibility, stationary, primal feasibility and complementary slackness conditions, respectively.

We decompose the dual feasibility condition~(\ref{eq:kkt_dual_feasibility}) into two cases: ${\mu(x)=0}$ and ${\mu(x) > 0}$.
For ${\mu(x) = 0}$, the stationary condition~(\ref{eq:kkt_stationary}) gives:
\begin{equation}
\Delta k(x) = 0,
\end{equation}
and with the primal feasibility condition~(\ref{eq:kkt_primal_feasibility}) this leads to:
\begin{equation}
\phi_{0}(x) \geq 0.
\end{equation}
For ${\mu(x) > 0}$, the complementary slackness condition~(\ref{eq:kkt_complementary_slackness}) implies:
\begin{equation}
\phi_{0}(x) + \phi_{1}(x) \Delta k(x) = 0.
\label{eq:kkt_complementary_slackness_mupos}
\end{equation}
Recall that ${\phi_{1}(x) \in \mathbb{R}^{n}}$ is a nonzero vector with right pseudoinverse ${\phi_{1}^+(x) = \phi_{1}^\top(x) / (\phi_{1}(x) \phi_{1}^\top(x))}$ and ${\phi_{0}(x) \in \mathbb{R}}$ is a scalar.
Then, we can express $\Delta k(x)$ from~(\ref{eq:kkt_complementary_slackness_mupos}) as:
\begin{equation}
\Delta k(x) = -\phi_{0}(x) \phi_{1}^+(x).
\end{equation}
Furthermore, we can show that ${\phi_{0}(x) < 0}$ holds by expressing $\phi_{0}(x)$ from~(\ref{eq:kkt_complementary_slackness_mupos}) and substituting the stationary condition~(\ref{eq:kkt_stationary}):
\begin{equation}
\phi_{0}(x) = -\phi_{1}(x) \Delta k(x) = -\mu(x) \phi_{1}(x) \phi_{1}^\top(x) < 0,
\end{equation}
where we used that ${\mu(x) > 0}$ and ${\phi_{1}(x) \phi_{1}^\top(x) > 0}$.

In summary, for ${\mu(x) = 0}$ we have ${\Delta k(x) = 0}$ and ${\phi_{0}(x) \geq 0}$, while ${\mu(x) > 0}$ implies ${\Delta k(x) = -\phi_{0}(x) \phi_{1}^+(x)}$ and ${\phi_{0}(x) < 0}$.
These can be written as:
\begin{equation}
\Delta k(x) =
\begin{cases}
0 & {\rm if}\ \phi_{0}(x) \geq 0, \\
-\phi_{0}(x) \phi_{1}^+(x) & {\rm if}\ \phi_{0}(x) < 0 ,
\end{cases}
\end{equation}
or more compactly as~\cite{ames2020safety}:
\begin{equation}
\Delta k(x) = \max\{ - \phi_{0}(x), 0 \} \phi_{1}^+(x).
\end{equation}
Since ${k(x) = k_{\rm d}(x) + \Delta k(x)}$, we finally obtain~(\ref{eq:QP_solution}) as the solution to the quadratic program~(\ref{eq:QP}).

\section{Technical Details of the Segway Application}
\label{sec:appdx_segway}

Here we derive the governing equations of the Segway model described in Section~\ref{sec:segway}, using Lagrange equations of the second kind.
This reproduces the model in~\cite{Gurriet2018a}.
Then, we describe the ECBF and the corresponding Lipschitz coefficients for the obstacle avoidance task.

\subsection{Segway Dynamics}
\label{sec:appdx_segway_dynamics}

The Segway's mechanical model is shown in Fig.~\ref{fig:segway}.
This planar model contains two rigid bodies: the frame and the wheels.
The two wheels are considered to be identical, hence they are treated together with their combined mass and inertia, while the voltage and torque at the two motors are assumed to be the same.
We denote the center of the wheels by point C, the center of mass (CoM) of the frame by point G, their distance by $L$ and the wheel radius by $R$.
We measure the pitch angle such that ${\varphi=0}$ in equilibrium, where G is located above C.
Note that the frame is asymmetric, and the frame axis is not vertical in equilibrium but it has an offset angle $\varphi_{0}$.

Assuming the wheels are rolling without slipping, the angular velocities $\omega_{\rm w}$ and $\omega_{\rm f}$ of the wheel and the frame and the velocities $v_{\rm C}$ and $v_{\rm G}$ of points C and G can be calculated by:
\begin{align}
\begin{split}
\omega_{\rm w} & = \dot{p}/R, \quad
\omega_{\rm f} = \dot{\varphi}, \\
v_{\rm C} & =
\begin{bmatrix}
\dot{p} \\ 0
\end{bmatrix}, \quad
v_{\rm G} =
\begin{bmatrix}
\dot{p} + L \dot{\varphi} \cos \varphi \\
- L \dot{\varphi} \sin \varphi
\end{bmatrix}.
\end{split}
\end{align}
Then, with the mass $M$ and mass moment of inertia $J_{\rm C}$ of the wheels
and the mass $m$ and mass moment of inertia $J_{\rm G}$ of the frame, the kinetic energy of the Segway is:
\begin{align}
\begin{split}
T & = \frac{1}{2} M v_{\rm C}^2 + \frac{1}{2} J_{\rm C} \omega_{\rm w}^2 + \frac{1}{2} m v_{\rm G}^2 + \frac{1}{2} J_{\rm G} \omega_{\rm f}^2 \\
& = \frac{1}{2} m_{0} \dot{p}^2 + m L \dot{p} \dot{\varphi} \cos \varphi + \frac{1}{2} J_{0} \dot{\varphi}^2,
\end{split}
\end{align}
where ${m_{0} = m + M + J_{\rm C}/R^2}$ and ${J_{0} = m L^2 + J_{\rm G}}$.
The potential energy of the Segway is:
\begin{equation}
U = m g L \cos \varphi.
\end{equation}
The power of the total driving torque $M_{\rm d}$ exerted by the two motors at the wheels can be expressed as:
\begin{equation}
P = M_{\rm d} ( \omega_{\rm w} - \omega_{\rm f} ) =  Q_{p} \dot{p} + Q_{\varphi} \dot{\varphi},
\end{equation}
yielding the general forces ${Q_{p} = M_{\rm d}/R}$ and ${Q_{\varphi} = - M_{\rm d}}$.

The driving torque $M_{\rm d}$ can be related to the voltage $u$ of the motors.
We regard the voltage as control input, obtained from the following motor model:
\begin{align}
\begin{split}
u & = R_{\rm a} i + K_{\rm b} (\omega_{\rm w} - \omega_{\rm f}), \\
M_{\rm d} & = K_{\rm t} i,
\end{split}
\end{align}
where $i$ is the armature current, $R_{\rm a}$ is the armature resistance, $K_{\rm b}$ is the back electromagnetic field constant and $K_{\rm t}$ is the torque constant of the motors.
This implies the driving torque:
\begin{equation}
M_{\rm d} = K_{\rm m} u - b_{\rm t} (\dot{p} - R \dot{\varphi}),
\end{equation}
with constants ${K_{\rm m} = K_{\rm t}/R_{\rm a}}$ and
${b_{\rm t} = K_{\rm t} K_{\rm b}/(R_{\rm a} R)}$.

With these preliminaries, we write Lagrange's equations:
\begin{align}
\begin{split}
\frac{{\rm d}}{{\rm d}t} \frac{\partial T}{\partial \dot{p}} - \frac{\partial T}{\partial p} + \frac{\partial U}{\partial p} = Q_{p}, \\
\frac{{\rm d}}{{\rm d}t} \frac{\partial T}{\partial \dot{\varphi}} - \frac{\partial T}{\partial \varphi} + \frac{\partial U}{\partial \varphi} = Q_{\varphi},
\end{split}
\end{align}
which, after substitution, lead to:
\begin{align}
\begin{split}
m_{0} \ddot{p} + m L \cos \varphi \ddot{\varphi} - m L \sin \varphi \dot{\varphi}^2 & = \frac{M_{\rm d}}{R}, \\
 m L \cos \varphi \ddot{p} + J_{0} \ddot{p} - m g L \sin \varphi & = - M_{\rm d}.
\end{split}
\end{align}
Ultimately, we obtain the equations of motion in the form:
\begin{equation}
D(q) \ddot{q} + H(q,\dot{q}) = B u,
\end{equation}
with the inertia matrix $D(q)$, Coriolis and gravity terms included in $H(q,\dot{q})$ and input matrix $B$:
\begin{align}
\begin{split}
D(q) & =
\begin{bmatrix}
m_{0} & m L \cos \varphi \\
m L \cos \varphi & J_{0}
\end{bmatrix}, \quad
B =
\begin{bmatrix}
K_{\rm m}/R \\ -K_{\rm m}
\end{bmatrix}, \\
H(q,\dot{q}) & =
\begin{bmatrix}
- m L \sin \varphi \dot{\varphi}^2 + b_{\rm t}/R (\dot{p} - R \dot{\varphi}) \\
- m g L \sin \varphi - b_{\rm t} (\dot{p} - R \dot{\varphi})
\end{bmatrix}.
\end{split}
\end{align}

The equations of motion can be rearranged to the first-order control-affine form~(\ref{eq:system}):
\begin{equation}
\begin{bmatrix}
\dot{q} \\ \ddot{q}
\end{bmatrix} =
\begin{bmatrix}
\dot{q} \\ - D^{-1}(q) H(q,\dot{q})
\end{bmatrix} +
\begin{bmatrix}
0 \\ D^{-1}(q) B
\end{bmatrix} u,
\end{equation}
which leads to:
\begin{equation}
\begin{bmatrix}
\dot{p} \\ \dot{\varphi} \\ \dot{v} \\ \dot{\omega}
\end{bmatrix} =
\begin{bmatrix}
v \\ \omega \\ f_{v}(\varphi,v,\omega) \\
f_{\omega}(\varphi,v,\omega)
\end{bmatrix} +
\begin{bmatrix}
0 \\ 0 \\ g_{v}(\varphi) \\ g_{\omega}(\varphi)
\end{bmatrix} u,
\end{equation}
cf.~(\ref{eq:segway_dynamics}).
The expressions of the drift terms are:
\begin{align}
\begin{split}
f_{v}(\varphi,v,\omega) & = \frac{a \omega^2 \sin \varphi - g \sin \varphi \cos \varphi}{b - \cos^2 \varphi}
- \kappa g_{v}(\varphi) (v - R \omega), \\
f_{\omega}(\varphi,v,\omega) & = \frac{c \sin \varphi - \omega^2 \sin \varphi \cos \varphi}{b - \cos^2 \varphi}
- \kappa g_{\omega}(\varphi) (v - R \omega),
\end{split}
\end{align}
whereas those of the control matrix read:
\begin{equation}
g_{v}(\varphi) = \frac{A + B \cos \varphi}{b - \cos^2 \varphi}, \quad
g_{\omega}(\varphi) = -\frac{C + D \cos \varphi}{b - \cos^2 \varphi},
\end{equation}
with parameters:
\begin{align}
\begin{split}
a & = \frac{J_{0}}{m L}, \quad
b = \frac{m_{0} J_{0}}{m^2 L^2}, \quad
c = \frac{m_{0} g}{m L}, \quad
\kappa = \frac{b_{\rm t}}{K_{\rm m}}, \\
A & = \frac{K_{\rm m} J_{0}}{m^2 L^2 R}, \;
B = \frac{K_{\rm m}}{m L}, \;
C = \frac{K_{\rm m} m_{0}}{m^2 L^2}, \;
D = \frac{K_{\rm m}}{m L R}.
\end{split}
\end{align}
The values of all parameters are listed in Table~\ref{tab:segway}.
These were identified for the Ninebot E+ Segway platform in~\cite{Gurriet2018a}.

\bgroup
\setlength{\tabcolsep}{3pt}
\begin{table}
\caption{Parameters of the Segway Model}
\label{tab:segway}
\begin{center}
\begin{tabular}{c c c c}
\hline
Description & Parameter & Value & Unit \\
\hline
gravitational acceleration & $g$ & 9.81 & m/s$^2$ \\
\hline
radius of wheels & $R$ & 0.195 & m \\
mass of wheels & $M$ & 2$\times$2.485 & kg \\
mass moment of inertia of wheels & $J_{\rm C}$ & 2$\times$0.0559 & kgm$^2$ \\
\hline
distance of wheel center and frame CoM  & $L$ & 0.169 & m \\
distance of wheel center and frame tip & $\ell$ & 0.75 & m  \\
mass of frame & $m$ & 44.798 & kg \\
mass moment of inertia of frame & $J_{\rm G}$ & 3.836 & kgm$^2$ \\
offset angle & $\varphi_{0}$ & 0.138 & rad \\
\hline
torque constant of motors & $K_{\rm m}$ & 2$\times$1.262 & Nm/V \\ 
damping constant of motors & $b_{\rm t}$ & 2$\times$1.225 & Ns \\
\hline
\multirow{10}{*}{combined parameters}
 & $m_{0}$ & 52.710 & kg \\
 & $J_{0}$ & 5.108 & kgm$^2$ \\
 & $a$ & 0.6768 & m \\
 & $b$ & 4.7274 & - \\
 & $c$ & 68.5205 & 1/s$^{2}$\\
 & $\kappa$ & 0.9713 & Vs/m \\
 & $A$ & 1.1605 & m/s$^2$/V \\
 & $B$ & 0.3344 & m/s$^2$/V \\
 & $C$ & 2.3355 & 1/s$^2$/V \\
 & $D$ & 1.7147 & 1/s$^2$/V \\
\hline
\end{tabular}
\end{center}
\end{table}
\egroup

\subsection{Expression of the ECBF and its Lipschitz Coefficients}
\label{sec:appdx_segway_Lipschitz}

Now we give the detailed expressions of the extended ECBF in~(\ref{eq:segway_ECBF_extension}) and the corresponding Lipschitz coefficients in~(\ref{eq:segway_robustify}).
The ECBF candidate in~(\ref{eq:segway_ECBF_candidate}) is of the form:
\begin{equation}
H(x,e) = h_0(x) + h_1(x)e + e^2,
\end{equation}
with coefficients:
\begin{align}
\begin{split}
h_0(x) & =\big(p  + \ell \sin (\varphi+\varphi_{0})\big)^2 \\
& \quad + \big(R+\ell \cos (\varphi+\varphi_{0}) - y\big)^2-r^2 , \\
h_1(x) & = -2\big(p + \ell \sin (\varphi+\varphi_{0})\big) .
\end{split}
\end{align}
Then, the extended ECBF in~(\ref{eq:segway_ECBF_extension}) becomes:
\begin{equation}
H_{\rm e}(x,e,\dot{e}) = H_0(x) + H_1(x)e + h_1(x)\dot{e} + \gamma_{\rm e}e^2+2e\dot{e},
\end{equation}
where:
\begin{align}
\begin{split}
H_0(x) & = \nabla_{p} h_0(x) v + \nabla_{\varphi} h_0(x) \omega + \gamma_{\rm e} h_0(x), \\
H_1(x) & = \nabla_{p} h_1(x) v + \nabla_{\varphi} h_1(x) \omega + \gamma_{\rm e} h_1(x).
\end{split}
\end{align}
Notice that $h_0$ and $h_1$ depend on the states $p$ and $\varphi$ only, whose derivatives are independent of the control input $u$.

The Lipschitz coefficients in~(\ref{eq:segway_robustify}) belong to the functions:
\begin{equation}
\begin{split}
\nabla_{x}H_{\rm e}f(x,e,\dot{e})&=C_0(x) + C_1(x)e + C_2(x)\dot{e},\\
\nabla_{x}H_{\rm e}g(x,e,\dot{e})&=C_3(x) + C_4(x)e,\\
\nabla_{e}H_{\rm e}\dot{e}(x,e,\dot{e})&=H_1(x)\dot{e} + 2\gamma_{\rm e}e\dot{e} + 2\dot{e}^2,\\
\nabla_{\dot{e}}H_{\rm e}\ddot{e}(x,e,\dot{e},\ddot{e})&=h_1(x)\ddot{e} + 2e\ddot{e}, \\
\alpha \circ H_{\rm e}(x,e,\dot{e}) &= \gamma H_0(x) + \gamma H_1(x)e + \gamma h_1(x)\dot{e} \\
& \quad + \gamma \gamma_{\rm e} e^2 + 2 \gamma e \dot{e},
\end{split}
\end{equation}
where:
\begin{align}
\begin{split}
C_0(x) & = \nabla_{x} H_0(x) f(x), \\
C_1(x) & = \nabla_{x} H_1(x) f(x), \\
C_2(x) & = \nabla_{x} h_1(x) f(x), \\
C_3(x) & = \nabla_{x} H_0(x) g(x), \\
C_4(x) & = \nabla_{x} H_1(x) g(x).
\end{split}
\end{align}

For example, to identify the Lipschitz coefficients of $\nabla_{e}H_{\rm e}\dot{e}$, we can write:
\begin{align}
\begin{split}
\nabla_{e}&H_{\rm e}\dot{e}(x,e,\dot{e}) - \nabla_{e}H_{\rm e}\dot{e}(x,\hat{e},\hat{\dot{e}}) \\
& = H_1(x) (e - \hat{e}) + 2 \gamma_{\rm e} (e\dot{e} - \hat{e}\hat{\dot{e}}) + 2 \big(\dot{e}^2 - \hat{\dot{e}}^2 \big) \\
& = H_1(x) (e - \hat{e}) + 2 \gamma_{\rm e} \dot{e} (e - \hat{e}) + 2 \gamma_{\rm e} \hat{e} (\dot{e} - \hat{\dot{e}}) \\
& \quad + 2 \hat{\dot{e}} (\dot{e} - \hat{\dot{e}}) + 2 \dot{e} (\dot{e} - \hat{\dot{e}}) \\
& \geq - (|H_1(x)| + 2 \gamma_{\rm e} \max_{\dot{e} \in D_{\dot{e}}} |\dot{e}|) |e - \hat{e}| \\
& \quad - (2 \gamma_{\rm e} |\hat{e}| + 2 |\hat{\dot{e}}| + 2 \max_{\dot{e} \in D_{\dot{e}}} |\dot{e}|) | \dot{e} - \hat{\dot{e}} |,
\end{split}
\end{align}
hence the corresponding Lipschitz coefficients are:
\begin{equation}
\begin{split}
\mathcal{L}_{\nabla H_{\rm e} \dot{e},e} & = |H_1(x)| + 2 \gamma_{\rm e}  \max_{\dot{e} \in D_{\dot{e}}} |\dot{e}| ,\\
\mathcal{L}_{\nabla H_{\rm e} \dot{e},\dot{e}} & = 2\gamma_{\rm e}|\hat{e}| + 2 |\hat{\dot{e}}| + 2 \max_{\dot{e} \in D_{\dot{e}}} |\dot{e}| .
\end{split}
\end{equation}
Here we considered that the unknown environment state derivative $\dot{e}$ is restricted to a domain ${D_{\dot{e}} \subseteq \mathcal{E}}$ to get local Lipschitz coefficients.
Similarly, the unknown environment state $e$ and acceleration $\ddot{e}$ can also be restricted to some domains ${D_{e} \subseteq E}$ and ${D_{\ddot{e}} \subseteq \mathbb{R}^{l}}$.
In the case of the Segway, we assumed that the obstacle's position and velocity are restricted to
${D_{e} = [-3 , 3]\,{\rm m}}$
and
${D_{\dot{e}} = [-0.55 , 0.55]\,{\rm m/s}}$
(while its acceleration was known to be zero).

After similar calculation, the list of the remaining Lipschitz coefficients is:
\begin{equation}
\begin{split}
\mathcal{L}_{\nabla H_{\rm e} f,e} & = |C_1(x)| ,\\
\mathcal{L}_{\nabla H_{\rm e} f,\dot{e}} & = |C_2(x)| , \\
\mathcal{L}_{\nabla H_{\rm e} g,e} & = |C_4(x)| ,\\
\mathcal{L}_{\nabla H_{\rm e} g,\dot{e}} & = 0 , \\
\mathcal{L}_{\nabla H_{\rm e} \ddot{e},e} & =  2  \max_{\ddot{e} \in D_{\ddot{e}}} |\ddot{e}| ,\\
\mathcal{L}_{\nabla H_{\rm e} \ddot{e},\dot{e}} & =0 ,\\
\mathcal{L}_{\nabla H_{\rm e} \ddot{e},\ddot{e}} & = |h_1(x)| + 2|\hat{e}| , \\
\mathcal{L}_{\alpha \circ H_{\rm e},e}  & = \gamma |H_1(x)| + \gamma \gamma_{\rm e}(|\hat{e}| + \max_{e \in D_{e}} |e|) + 2\gamma \max_{\dot{e} \in D_{\dot{e}}} |\dot{e}| , \\
\mathcal{L}_{\alpha \circ H_{\rm e},\dot{e}}  & = \gamma |h_1(x)|  + 2\gamma |\hat{e}|.
\end{split}
\end{equation}
Note that these coefficients may depend on the state $x$ or the estimates $\hat{e}$, $\hat{\dot{e}}$ and $\hat{\ddot{e}}$ to reduce conservatism, while they are independent of the unknown values $e$, $\dot{e}$ and $\ddot{e}$.




\bibliographystyle{IEEEtran}
\bibliography{2021_tcst}

\vspace{-30pt}
\begin{IEEEbiography}
[{\includegraphics[width=1in,height=1.25in,clip,keepaspectratio]{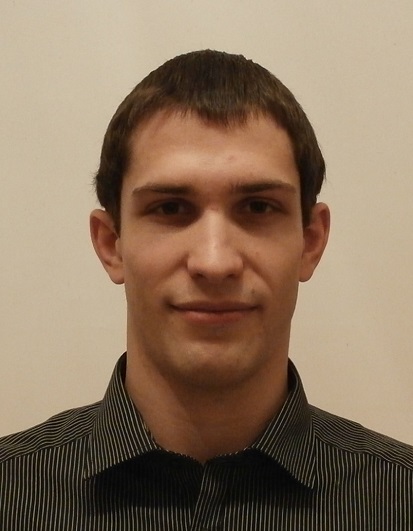}}]
{Tamas G. Molnar} received his B.Sc. degree in Mechatronics Engineering, M.Sc. and Ph.D. degrees in Mechanical Engineering from the Budapest University of Technology and Economics, Hungary, in 2013, 2015 and 2018.
He held postdoctoral position at the University of Michigan, Ann Arbor between 2018 and 2020.
Since 2020 he is a postdoctoral fellow at the California Institute of Technology, Pasadena.
His research interests include nonlinear dynamics and control, safety-critical control, and time delay systems with applications to connected automated vehicles, robotic systems, and machine tool vibrations.
\end{IEEEbiography}

\vspace{-30pt}
\begin{IEEEbiography}
[{\includegraphics[width=1in,height=1.25in,clip,keepaspectratio]{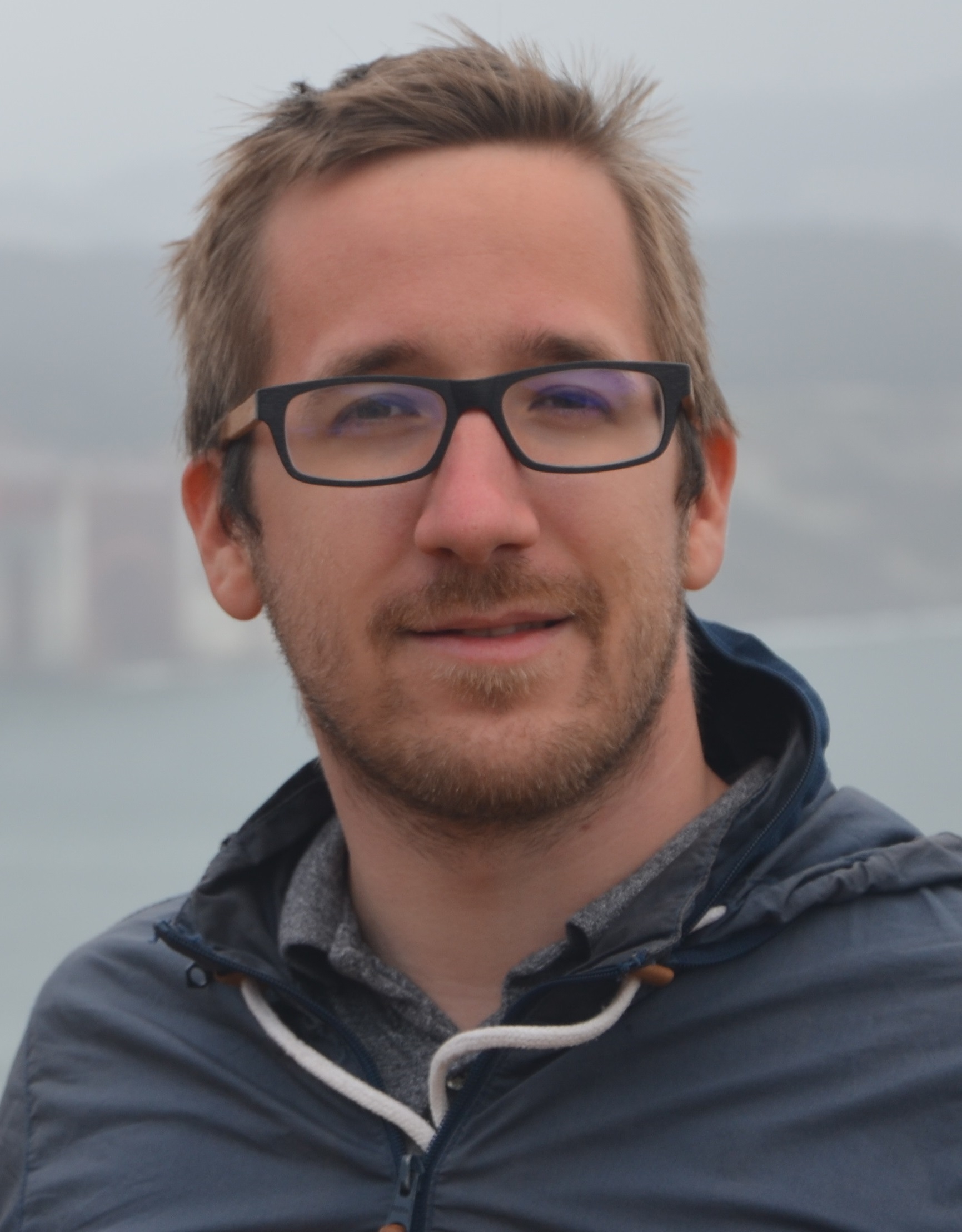}}]
{Adam K. Kiss} received his B.Sc. and M.Sc. degrees in mechanical engineering from the Budapest University of Technology and Economics (BME) in 2013 and 2015, where he is currently pursuing the Ph.D. degree.
He is currently a research assistant at the MTA-BME Lend{\"{u}}let Machine Tool Vibration Research Group and at the Department of Applied Mechanics, BME. His current research interests include nonlinear dynamics, safety-critical control and time delay systems with applications to machine tool vibrations and connected automated vehicles.
\end{IEEEbiography}

\vspace{-30pt}
\begin{IEEEbiography}
[{\includegraphics[width=1in,height=1.25in,clip,keepaspectratio]{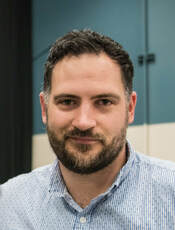}}]
{Aaron D. Ames} is the Bren Professor of Mechanical and Civil Engineering and Control and Dynamical Systems at Caltech. Prior to joining Caltech in 2017, he was an Associate Professor at Georgia Tech in the Woodruff School of Mechanical Engineering and the School of Electrical {\&} Computer Engineering. He received a B.S. in Mechanical Engineering and a B.A. in Mathematics from the University of St. Thomas in 2001, and he received a M.A. in Mathematics and a Ph.D. in Electrical Engineering and Computer Sciences from UC Berkeley in 2006. He served as a Postdoctoral Scholar in Control and Dynamical Systems at Caltech from 2006 to 2008, and began his faculty career at Texas A{\&}M University in 2008. At UC Berkeley, he was the recipient of the 2005 Leon O. Chua Award for achievement in nonlinear science and the 2006 Bernard Friedman Memorial Prize in Applied Mathematics, and he received the NSF CAREER award in 2010, the 2015 Donald P. Eckman Award, and the 2019 IEEE CSS Antonio Ruberti Young Researcher Prize.  His research interests span the areas of robotics, nonlinear, safety-critical control and hybrid systems, with a special focus on applications to dynamic robots -— both formally and through experimental validation. 
\end{IEEEbiography}

\vspace{-30pt}
\begin{IEEEbiography}
[{\includegraphics[width=1in,height=1.25in,clip,keepaspectratio]{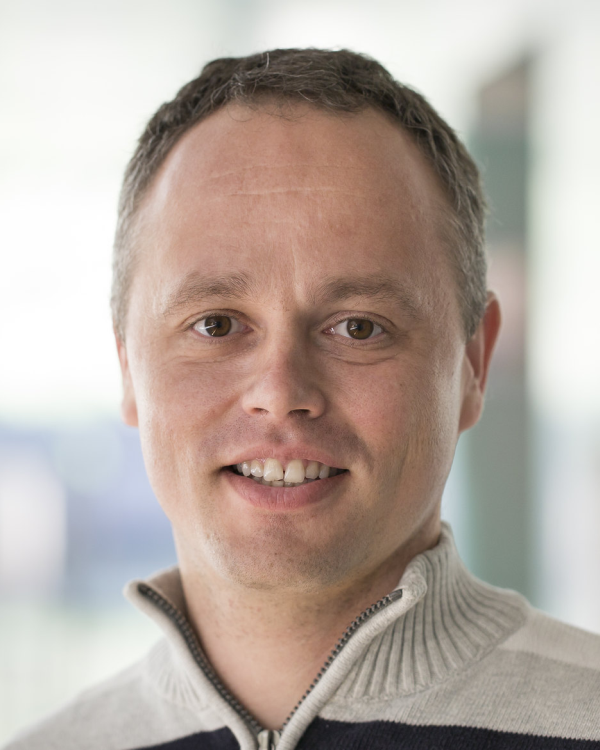}}]
{G{\'{a}}bor Orosz} received the M.Sc. degree in Engineering Physics from the Budapest University of Technology, Hungary, in 2002 and the Ph.D. degree in Engineering Mathematics from University of Bristol, UK, in 2006. He held postdoctoral positions at the University of Exeter, UK, and at the University of California, Santa Barbara. In 2010, he joined the University of Michigan, Ann Arbor where he is currently an Associate Professor in Mechanical Engineering and in Civil and Environmental Engineering. His research interests include nonlinear dynamics and control, time delay systems, and machine learning with applications to connected and automated vehicles, traffic flow, and biological networks.
\end{IEEEbiography}

\end{document}